\documentclass[12pt]{article}
\pdfoutput=1
\usepackage{jheppub}
\usepackage{amsmath}
\usepackage{amsfonts}
\usepackage{amssymb}
\usepackage{graphicx}

\usepackage[export]{adjustbox}
\newcommand{\bmat}{\left(\begin{array}}
\newcommand{\emat}{\end{array}\right)}

\def\yzero{\smash{\hbox{$y\kern-4pt\raise1pt\hbox{${}^\circ$}$}}}

\def\d{\delta}
\def\beq{\begin{equation}}
\def\eeq{\end{equation}}
\def\beqa{\begin{eqnarray}}
\def\eeqa{\end{eqnarray}}

\def\-{\hphantom{-}}

\def\s2{\frac{1}{\sqrt2}}

\def\beq{\begin{equation}}
\def\eeq{\end{equation}}
\def\beqa{\begin{eqnarray}}
\def\eeqa{\end{eqnarray}}

\def\IF{\relax{\rm I\kern-.18em F}}
\def\II{\relax{\rm I\kern-.18em I}}

\def\Dsl{\,\raise.15ex\hbox{/}\mkern-13.5mu D} 

\def\IS{{\bf {S}}}
\def\IR{{\bf {R}}}

\def\NN{{\cal {N}}}

\catcode`\@=11   
\newdimen\@rotdimen
\newbox\@rotbox  

\def\@vspec#1{\special{ps:#1}}
\def\@rotstart#1{\@vspec{gsave currentpoint currentpoint translate
   #1 neg exch neg exch translate}}
\def\@rotfinish{\@vspec{currentpoint grestore moveto}}
\def\@rotr#1{\@rotdimen=\ht#1\advance\@rotdimen by\dp#1%
   \hbox to\@rotdimen{\hskip\ht#1\vbox to\wd#1{\@rotstart{90 rotate}%
   \box#1\vss}\hss}\@rotfinish}
\def\@rotl#1{\@rotdimen=\ht#1\advance\@rotdimen by\dp#1%
   \hbox to\@rotdimen{\vbox to\wd#1{\vskip\wd#1\@rotstart{270 rotate}%
   \box#1\vss}\hss}\@rotfinish}%
\def\@rotu#1{\@rotdimen=\ht#1\advance\@rotdimen by\dp#1%
   \hbox to\wd#1{\hskip\wd#1\vbox to\@rotdimen{\vskip\@rotdimen
   \@rotstart{-1 dup scale}\box#1\vss}\hss}\@rotfinish}%
\def\@rotf#1{\hbox to\wd#1{\hskip\wd#1\@rotstart{-1 1 scale}%
   \box#1\hss}\@rotfinish}%
\def\rotate{\@ifnextchar[{\@rotate}{\@rotate[l]}}
\def\@rotate[#1]#2{\setbox\@rotbox=\hbox{#2}\@nameuse{@rot#1}\@rotbox}

\catcode`\@=12

\begin{document}

\makeatletter
\@addtoreset{equation}{section}
\makeatother
\renewcommand{\theequation}{\thesection.\arabic{equation}}
\pagestyle{empty}
\vspace*{0.5in}
\rightline{IFT-UAM/CSIC-23-68}
\vspace{1.5cm}
\begin{center}
\Large{\bf Intersecting End of the World Branes
}
\\[8mm] 

\large{Roberta Angius, Andriana Makridou,  Angel M. Uranga\\[4mm]}
\footnotesize{Instituto de F\'{\i}sica Te\'orica IFT-UAM/CSIC,\\[-0.3em] 
C/ Nicol\'as Cabrera 13-15, 
Campus de Cantoblanco, 28049 Madrid, Spain}\\ 
\footnotesize{\href{roberta.angius@csic.es}{roberta.angius@csic.es},  \href{mailto: andriana.makridou@ift.csic.es}{andriana.makridou@ift.csic.es},  \href{mailto:angel.uranga@csic.es}{angel.uranga@csic.es}}

\vspace*{10mm}

\small{\bf Abstract} \\
\end{center}
\begin{center}
\begin{minipage}[h]{\textwidth}

Dynamical cobordisms implement the swampland cobordism conjecture in the framework of effective field theory, realizing codimension-1 end of the world (ETW) branes as singularities at finite spacetime distance at which scalars diverge to infinite field space distance. ETW brane solutions provide a useful probe of infinity in moduli/field spaces and the associated swampland constraints, such as the distance conjecture.

We construct explicit solutions describing intersecting ETW branes in theories with multiple scalars and general potentials, so that different infinite field space limits coexist in the same spacetime, and can be simultaneously probed by paths approaching the ETW brane intersection. 
Our class of solutions includes physically interesting examples, such as intersections of Witten’s bubbles of nothing in toroidal compactifications, generalizations in compactifications on products of spheres, and possible flux dressings thereof (hence including charged objects at the ETW branes).
From the cobordism perspective, the intersections can be regarded as describing the end of the world for end of the world branes, or as boundary domain walls interpolating between different ETW brane boundary conditions for the same bulk theory. 
\end{minipage}
\end{center}
\newpage
\setcounter{page}{1}
\pagestyle{plain}
\renewcommand{\thefootnote}{\arabic{footnote}}
\setcounter{footnote}{0}

\tableofcontents

\vspace*{1cm}

\newpage

\section{Introduction}

One of the main outcomes of the swampland program (see \cite{Palti:2019pca,vanBeest:2021lhn,Grana:2021zvf,Agmon:2022thq} for reviews) is a renewed interest in the exploration of regions at infinite distance in moduli space.
A prominent tool and motivation is the Distance Conjecture \cite{Ooguri:2006in}, which posits the existence of towers of particles becoming exponentially light along trajectories reaching such infinite distance regions.

In theories with exact moduli spaces, such as the much studied case of 4d $\NN=2$ supersymmetry, the exploration of infinite distances is possible using spacetime independent scalar vevs. In this context there is a rich industry of various approaches and results, including \cite{Grimm:2018ohb,Lee:2018urn,Grimm:2018cpv,Corvilain:2018lgw,Marchesano:2019ifh,Lee:2019xtm,Lee:2019wij,Baume:2019sry,Gendler:2020dfp,Calderon-Infante:2020dhm,Etheredge:2022opl,Etheredge:2023odp,Castellano:2023stg,Castellano:2023jjt}. 
An interesting feature in theories with several scalars, in particular in CY moduli spaces \cite{Grimm:2018ohb,Grimm:2018cpv,Corvilain:2018lgw}, is the existence of a rich network of infinite distance loci with different components which in general intersect in non-trivial ways, and for which the Distance Conjecture requires formulations including the interplay of multiple towers \cite{Calderon-Infante:2020dhm,Etheredge:2022opl,Etheredge:2023odp,Castellano:2023stg,Castellano:2023jjt}.

In the presence of general scalar potentials, however, the above adiabatic exploration of infinite distances by constant vevs may result in inconsistencies \cite{Mininno:2020sdb} and may even be forbidden \cite{Gonzalo:2018guu} (see \cite{Demulder:2023vlo} for a recent discussion). One is thus bound to the study of spacetime-dependent solutions, as pioneered in \cite{Buratti:2018xjt} (see \cite{Etheredge:2023odp} for recent discussions, and \cite{Rudelius:2021azq,Calderon-Infante:2022nxb} for recent time-dependent running solutions and the Distance Conjecture). In this context, there are several classes of solutions describing scalars running to infinite field space distance at finite spacetime distance. These include dynamical cobordisms \cite{Buratti:2021yia,Buratti:2021fiv,Angius:2022aeq,Blumenhagen:2022mqw,Blumenhagen:2023abk} (see also 
\cite{Dudas:2000ff,Blumenhagen:2000dc,Dudas:2002dg,Dudas:2004nd,Hellerman:2006nx,Hellerman:2006ff,Hellerman:2007fc} for related early work and \cite{Basile:2018irz,Antonelli:2019nar,Mininno:2020sdb,Basile:2020xwi,Mourad:2021qwf,Mourad:2021roa,Basile:2021mkd,Mourad:2022loy,Angius:2022mgh,Basile:2022ypo,Angius:2023xtu,Huertas:2023syg} for other related recent developments), 4d EFT strings \cite{Lanza:2020qmt,Lanza:2021udy,Marchesano:2022avb}, and small black holes \cite{Hamada:2021yxy,Angius:2022aeq,Angius:2023xtu,Calderon-Infante:2023uhz} (see \cite{Delgado:2022dkz} for the exploration of infinity in moduli space using {\em large} black holes). 

Dynamical cobordisms describe configurations of scalars running in one spacetime dimension, 
along which spacetime ends at finite distance when the theory hits a spacetime singularity at 
which scalars run off to infinite field space distance \cite{Buratti:2021yia,Buratti:2021fiv}. They can be regarded as describing boundaries of spacetime at a codimension-1 end of the world (ETW) brane, which provide a dynamical realization of the cobordism defects predicted by the Cobordism Conjecture of \cite{McNamara:2019rup} (see \cite{GarciaEtxebarria:2020xsr,Ooguri:2020sua,Montero:2020icj,Dierigl:2020lai,Hamada:2021bbz,Blumenhagen:2021nmi,Blumenhagen:2022bvh,Dierigl:2022reg,Debray:2023yrs,Dierigl:2023jdp,Basile:2023knk,Friedrich:2023tid} for other applications). Interestingly, they admit a universal local description introduced in \cite{Angius:2022aeq} in terms of single parameter (dubbed critical exponent), which moreover controls interesting scaling relations between the spacetime and field theory distances in the solution.

A natural question is how to use running solutions to explore the network of infinite distance limits in theories with multiple scalars. To achieve this goal, we consider the generalization of the above configurations, by considering solutions which include different spacetime regions at which different infinite distance limits are attained, and which intersect in spacetime so as to allow the exploration of the intersection of components of the infinite distance loci in field space. In particular we focus on the realization of this idea using dynamical cobordisms as building blocks, and build a large class of explicit solutions describing intersecting ETW brane configurations. 

\begin{figure}[htb]
\begin{center}
\includegraphics[scale=.37]{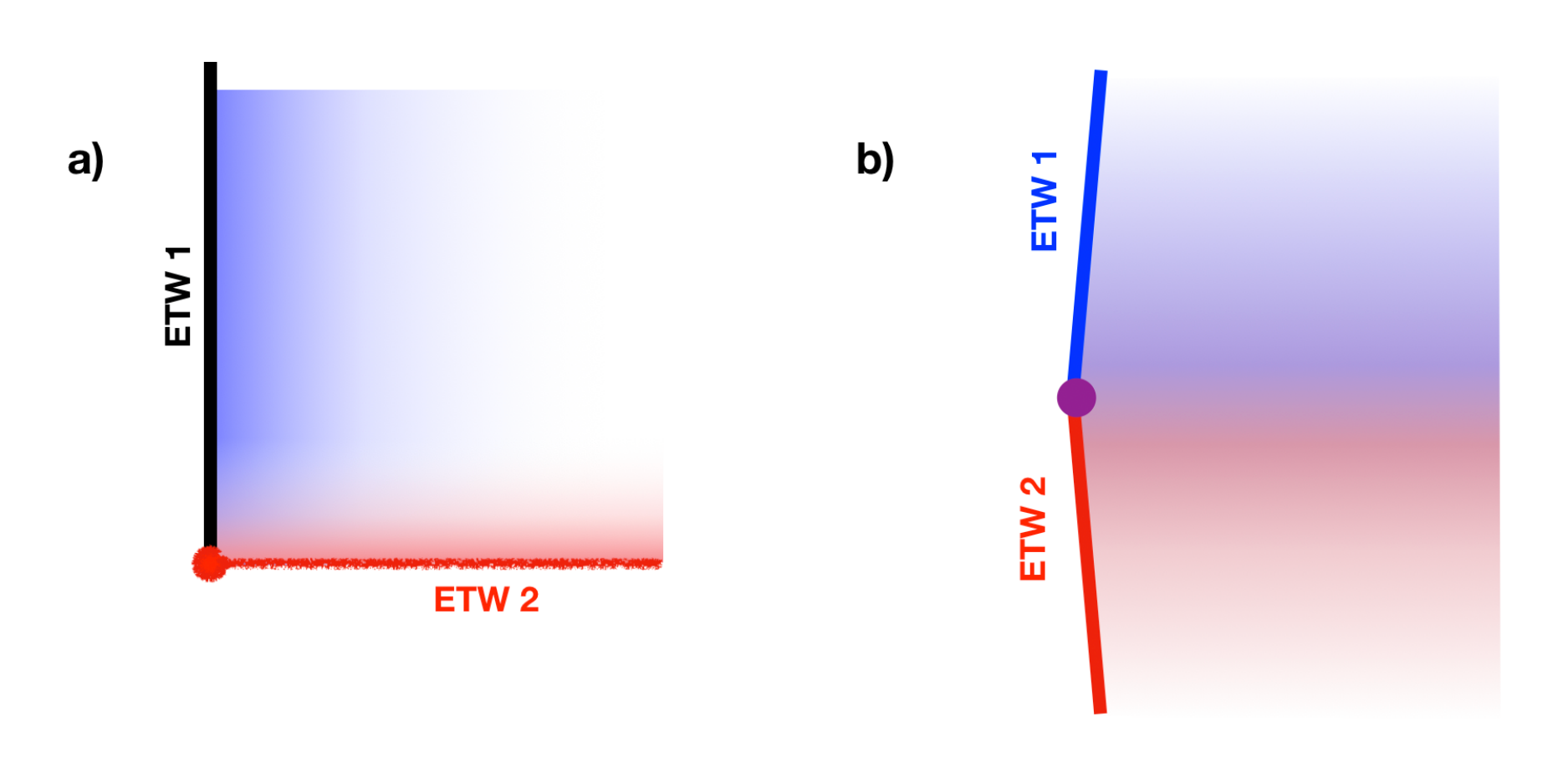}
\label{fig:ETW-of-ETW}
\caption{\small Two possible cobordism interpretations of intersecting ETW branes: a) The ETW$_2$ brane defines a cobordism to nothing for the configuration of the bulk theory ending at the ETW$_1$ brane boundary. b)The bulk theory ends on a cobordism to nothing boundary which switches from and ETW$_1$ brane to and ETW$_2$ brane.}
\end{center}
\end{figure}

Intersecting ETW branes have further interesting interpretations in the light of other swampland conjectures besides the Distance Conjecture. Being a key ingredient in dynamical cobordism, they have a natural home in the Cobordism Conjecture \cite{McNamara:2019rup}. Indeed, a configuration of two intersecting ETW branes (ETW$_1$ and ETW$_2$) can be regarded as a dynamical cobordism where the ETW$_2$ brane defines a boundary for the configuration of the bulk theory ending on the ETW$_1$ (and viceversa), see figure \ref{fig:ETW-of-ETW}a. In short, the intersection provides the end of the world for end of the world branes.

A second cobordism interpretation for the intersecting ETW brane configurations, illustrated in figure \ref{fig:ETW-of-ETW}b, is as providing a domain wall between different boundaries, defined by the ETW$_1$ and ETW$_2$ branes, for the same bulk theory. This is again in the spirit of the Cobordism Conjecture, which implies that in quantum gravity theories there must exist domain walls interpolating between any two configurations.

\medskip

The full expicit solutions we construct have a remarkably simple structure, realizing a superposition of the individual ETW branes, and are fully characterized by the individual critical exponents. In fact we compute the source terms  and find they correspond to localized terms associated to the ETW brane tension and scalar couplings, with no additional source terms localized at their intersection. 

Our approach should be regarded as local near the intersection, in the spirit of \cite{Angius:2022aeq}, and similarly leads to a universal scaling properties. We explore them in the context of versions of the Distance Conjecture in theories with multiple scalars, including the Convex Hull formulation \cite{Calderon-Infante:2020dhm}.

Although we do not expect our solutions to provide the most general intersecting ETW brane solutions, we show that they include  a rich set of physically relevant systems, such as intersecting Witten’s bubbles of nothing \cite{Witten:1981gj} (see \cite{Blanco-Pillado:2023aom,Blanco-Pillado:2023hog,Sugimoto:2023oul,Delgado:2023uqk} for recent related systems) in toroidal compactifications, or generalizations in compactifications in products of spheres, possibly dressed with fluxes (and hence including D-brane sources at the ETW brane). The setup also provides an arena for the actual exploration of the network of infinite distance loci in CY moduli spaces, which will be discussed in \cite{Angius:2024}.

For simplicity we mostly focus on configurations of two ETW branes intersecting orthogonally (see \cite{Blumenhagen:2000dc,Dudas:2002dg,Blumenhagen:2022mqw,Blumenhagen:2023abk} for other discussions of codimension-2 solutions), but also present generalizations for more than two ETW branes, and for general angles. We also compare our solutions with intersecting ETW branes associated to the {\em same} scalar, and show the latter actually are better regarded as singular limits of a single recombined ETW brane. This fits with the interpretation in \cite{Angius:2022mgh} for a particular example with tachyon condensation in supercritical strings.

A cosmological application of our solutions would be the description of collisions of cosmological bubbles (see \cite{Kleban:2011pg} for a review). It would be interesting to exploit our local models to extract universal signatures of these phenomena. We leave the exploration of phenomenological applications of our solutions to future work.

The paper is organized as follows. In section \ref{sec:codim1} we review the codimension-1 ETW brane solution, following \cite{Angius:2022aeq}. In section \ref{sec:intersecting} we construct the intersecting ETW brane solutions and discuss their properties. The ansatz and explicit solutions are constructed in section \ref{sec:ansatz}, section \ref{sec:sources} discusses the associated ETW brane worldvolume source terms, and in section \ref{sec:scalings} we describe the scaling properties of the class of solutions. Section \ref{sec:examples} contains explicit examples, including intersections of Witten's bubbles of nothing in $\IS^1\times\IS^1$ compactifications (section \ref{example-s1s1}), and in $\IS^{p_1}\times\IS^{p_2}$ compactifications (section \ref{sec:sp1sp2}), ETW branes with charged D-brane defects (section \ref{sec:dbrane}), and the intersection of a bubble of nothing with a general ETW brane (section \ref{sec:one-general-ETW}). In section \ref{sec:swampland} we discuss the interplay with swampland constraints, including the cobordism conjecture (section \ref{sec:swamp-cobordism}), the distance conjecture (section \ref{sec:swamp-distance}), including the convex hull formulation in \cite{Calderon-Infante:2020dhm} and the infinite distance pattern in \cite{Castellano:2023stg,Castellano:2023jjt}. Finally, we offer some concluding remarks in section \ref{sec:conclu}.
Appendix \ref{sec:generalizations} discusses several generalizations and related systems, including intersections at general angles (section \ref{sec:angles}), intersections of more than two ETW branes (section \ref{sec:triple}), and intersecting ETW branes with a single scalar (section \ref{sec:single})

\section{Overview of codimension-1 ETW branes}
\label{sec:codim1}

In this section we overview the local description for codimension-1 ETW branes in \cite{Angius:2022aeq}. Consider $d$-dimensional gravity coupled to a real scalar $\phi$ with general potential
\begin{equation}
	S = \int d^{d}x\, \sqrt{-g}\,  \left( \frac{1}{2}R - \frac{1}{2} \left( \partial \phi\right)^{2} - V(\phi) \right) \, , 
	\label{ddim-action}
\end{equation}
Here and in the rest of this work we set $M_{Pl}=1$ units and consider $d>2$. The scalar may correspond to a combination of underlying moduli/scalar fields.

The codimension-1 ETW brane solution has the structure
\beqa 
ds^{2}_d & = &e^{-2\sigma(y)} ds_{d-1}^{2} + dy^{2} \, , \nonumber\\
\phi & = &\phi(y) \, , 
\label{dw-ansatz}
\eeqa
As pioneered in \cite{Buratti:2021fiv}, the dynamical cobordism is characterized by the scalar $\phi$ going off to infinite distance in field space $\phi\to\infty$ at finite distance in spacetime $y\to 0$. Imposing the equations of motion, the local description near the ETW brane is
\beqa
\phi(y) &\simeq &- \frac{2}{\delta} \log y \nonumber\\
	\sigma(y) &\simeq &- \frac{4}{(d-2)\delta^{2}} \,\log y \, + \frac{1}{2} \log c\, ,
 \label{local-codim1-etw}
\eeqa
where the parameter $\delta$ describes the leading exponential behaviour of the potential
\beqa
V (\varphi) = -ac e^{\delta \varphi}
\label{codim1-potential}
\eeqa 
with $c$ a free parameter and $a$ is related to $\delta$ by
\beqa
\delta=\sqrt{\frac{d-1}{d-2}(1-a)}
\eeqa

As explained in \cite{Angius:2022aeq} the above solution leads to universal scaling relations among the spacetime distance to the singularity $\Delta$, the traverse scalar field space distance ${\cal D}$ and the spacetime curvature scalar, in terms of the parameter $\delta$:
\beqa
\Delta \sim e^{-\frac 12\delta {\cal D}}, \quad |R|\sim e^{\delta {\cal D}}
\label{universal-scalings-codim1}.
\eeqa 

A general warning, for these solutions and those in coming sections, is that at infinity in field space one expects the theory to have a lowered cutoff, which limits the validity of the effective field theory. Indeed, the appearance of corrections at the species scale has been discussed in small black holes in \cite{Calderon-Infante:2023uhz}, and we may expect similar phenomena in ETW brane solutions. Still, effective field theory remains a useful tool to describe the systems, and even to quantify these corrections. Our solutions in this work should be understood in this spirit.

\medskip

We now discuss a simple example of ETW brane, given by an analogue of the bubble of nothing of $\IS^1$ compactifications \cite{Witten:1981gj}, with the role of the expanding bubble replaced by a flat static wall of nothing (see \cite{Angius:2022aeq} for the spherical bubble case).

We start with $(d+1)$-dimensional gravity with an action
\beqa
S_{d+1}=\frac 12\int d^{d+1} x \sqrt{-g}R .
\eeqa
We consider compactifying on $\IS^1$ parametrized by $\theta$, with the compactification ansatz
\beqa
ds_{d+1}^2= e^{\alpha\rho}ds_{d}^2 + e^{-\beta \rho} d\theta^2.
\label{ansatz-codim1}
\eeqa
The parameters $\alpha$, $\beta$ are fixed by requiring the $d$-dimensional metric is in the Einstein frame, and by fixing the normalization of the radion kinetic term. We have
\beqa
 \beta = (d-2) \alpha \quad ,\quad 
     \alpha^2 = \frac{4}{(d-1)(d-2)} \, .
\eeqa
\noindent
The resulting $d$-dimensional action is
\beqa
S=\frac 12\int d^{d} x \, \sqrt{-g}\,  \big[ R -  ( \partial \rho)^{2}\big]\, .
\eeqa
The theory is like (\ref{ddim-action}) with the scalar $\phi=\rho$ and zero potential, since $\IS^1$ has no curvature. It thus admits a solution of the kind (\ref{dw-ansatz}), (\ref{local-codim1-etw}) with
\beqa
\delta= 2 \sqrt{\frac{d-1}{d-2}}.
\eeqa
It is easy to uplift this solution and check that it corresponds to taking an $\IS^1$ slicing of $(d+1)$ dimensional flat space
\beqa
ds_{d+1}^2=ds_{d-1}^2+dr^2+r^2d\theta^2,
\label{flat-s1-sliced}
\eeqa
with $ds_{d-1}^2$ describing a flat metric along the ETW brane worldvolume, and $r$ and $y$ related by
\beqa
y= \left( \frac{d-2}{d-1} \right) r^{\frac{d-1}{d-2}}.
\eeqa

We refer to \cite{Angius:2022aeq} for other examples, some of which will be arise as building blocks in our examples of intersecting ETW branes in section \ref{sec:examples}. We now turn to the general description of codimension-2 intersections of ETW branes.

\section{Intersecting ETW branes}
\label{sec:intersecting}

In this section we consider configurations describing the intersection of two ETW branes of the kind considered in the previous section. As we will see, they remarkably satisfy a simple superposition ansatz. This is reminiscent of the superposition of harmonic functions for supergravity solutions of (suitably smeared) intersecting BPS branes, with the differences that we do not require supersymmetry of the solutions, or even of the underlying theory, and that our solutions are fully localized and require no smearing.

We note that we focus on solutions describing the local behaviour near the intersection. The global structure in a general setup may differ in a model-dependent way. Hence, we focus on the universal behaviour of the configurations, much in the spirit of \cite{Angius:2022aeq} for the codimension-1 case.

\subsection{Codimension-2 ansatz and solutions}
\label{sec:ansatz}

We consider the following $(n+2)$-dimensional action for gravity coupled to two real scalars with a general potential $V (\phi_1 , \phi_2)$:
\begin{equation}
    S= \int d^{n+2}x \sqrt{-g} \left\lbrace \frac{1}{2} R - \frac{1}{2} \left( \partial \phi_1 \right)^2 - \frac{1}{2} \left( \partial \phi_2 \right)^2 - \frac{\alpha}{2} \partial_{\rho} \phi_1 \partial^{\rho} \phi_2- V (\phi_1 , \phi_2) \right\rbrace.
    \label{action_2scalars}
\end{equation}
Note that we have introduced a mixed kinetic term, which could be removed by diagonalization. However, maintaining it will allow for a simpler solution for the scalar profiles. As in the codimension-1 case in section \ref{sec:codim1}, the scalars can be combinations of several moduli/scalar fields.

The above action is regarded as describing the theory around the infinite distance locus. An important observation in this respect is that we are using a locally flat metric around that point. Points at infinity are actually singular  in general, but can admit such description if one restricts to specific directions in field space. An illustrative example is provided by considering two complex scalars $\Phi$ with K\"ahler potential
\beqa
K=\log(\Phi_1+{\overline\Phi}_1)+\log(\Phi_1+{\overline\Phi}_2) .
\eeqa
Introducing the axion and saxion components $\Phi_i=\varphi_i+it_i$, the metric is given by two decoupled hyperbolic spaces
\beqa
\frac{1}{t_1^2}\left(\,dt_1^2+d\varphi_1^2\,\right)+
\frac 1{t_2^2} \left(\,dt_2^2+d\varphi_2^2\,\right)= d\phi_1^2+d\phi_2^2+e^{-2\phi_1} d\varphi_1^2+e^{-2\phi_2} d\varphi_2^2 ,
\eeqa
where we have introduced the canonically normalized saxion fields $\phi_i=\log t_i$. Clearly, the metric for the axions $\varphi_i$ is singular at the infinite distance locus for $\phi_i\to\infty$. On the other hand, restricting to solutions where the axions are inactive, the dynamics for the saxions is controlled by flat metric kinetic terms. The action (\ref{action_2scalars}) should be understood as describing such smooth slices around the infinite distance point \footnote{More formally, in the formalism of \cite{Calderon-Infante:2020dhm}, the smooth slice belongs to the subspace $\mathbb{G}$ spanned by asymptotic tangent vectors of asymptotically geodesic trajectories.}.

In order to solve the equations of motion for the above action, we consider the ansatz for the metric:
\begin{equation}
    ds^2_{n+2} = e^{2 A(y_1, y_2)} ds^2_n + e^{2 B(y_1, y_2)} dy_1^2 + e^{2 C(y_1,y_2)} dy_2^2\, ,
\label{codim2-ansatz}    
\end{equation}
and for the profiles of the two scalars:
\begin{equation}
    \phi_1 = \phi_1 (y_1)\quad, \quad \quad \quad \phi_2 = \phi_2 (y_2).
\end{equation}
A linear independent set of combinations of the equations of motion is 
\begin{equation}
    \begin{split}
         (1) \quad & e^{-2C}n \left[ n \left( A' \right)^2 +  A'  \left( B'-C' \right) + A'' \right] +  e^{-2B} n\left[ n \left(\dot{A} \right)^2 + \dot{A}  \left(\dot{C} - \dot{B} \right) + \ddot{A} \right] =-2V , \\
         (2) \quad & e^{-2C} \left[ -n \left( A' \right)^2 +(n-2)  A' \left( B'-C' \right) +(n-2)  A''  +2 B'  \left(B'-C' \right) +2  B'' + \left(  \phi_2' \right)^2 \right] +\\
          + & e^{-2B} \left[ -n \left( \dot{A} \right)^2 +(n-2) \dot{A}  \left( \dot{C}-\dot{B}\right) +(n-2) \ddot{A} +2 \dot{C}  \left(\dot{C}-\dot{B} \right) +2 \ddot{C} + \left( \dot{ \phi_1 }\right)^2 \right] =0 , \\
         (3) \quad & n \left[\dot{A} B' - \dot{A} A' + \dot{C} B' - \dot{A}' \right] = \frac{\alpha}{2} \dot{\phi_1}  \phi_2' , \\
         (4) \quad & e^{-2B} \left[ \left( n \dot{A}-\dot{B}+\dot{C} \right) \dot{\phi_1} + \ddot{\phi_1} \right] + \frac{\alpha}{2} e^{-2C} \left[ \left( nA' +B'-C' \right) \phi_2' +  \phi_2'' \right] - \frac{\partial V}{\partial \phi_1} =0 ,  \\
         (5) \quad & e^{-2C} \left[ \left(nA' +B'-C' \right) \phi_2' + \phi_2'' \right]  +  \frac{\alpha}{2} e^{-2B} \left[ \left( n\dot{A}-\dot{B}+\dot{C} \right) \dot{\phi_1} +  \ddot{\phi_1} \right] - \frac{\partial V}{\partial \phi_2}=0 ,
    \end{split}
    \label{lin_indep_eoms}
\end{equation}
where we have introduced the notation $\dot{f}\equiv\partial_{y_1}f$, $f'\equiv \partial_{y_2}f$. \\
In particular, eq.$(1)$ is proportional to the sum of the  $\{y_1 y_1\}$ and $\{y_2 y_2\}$ components of the Einstein equations. Eq.$(2)$ is the $\{ij\}$ component, using eq.$(1)$ to eliminate the potential. Eq.$(3)$ is the mixed $\{y_1 y_2\}$component, and eqs.$(4)$ and $(5)$ are the equations of motion for the scalars.

\begin{figure}[htb]
\begin{center}
\includegraphics[scale=.18]{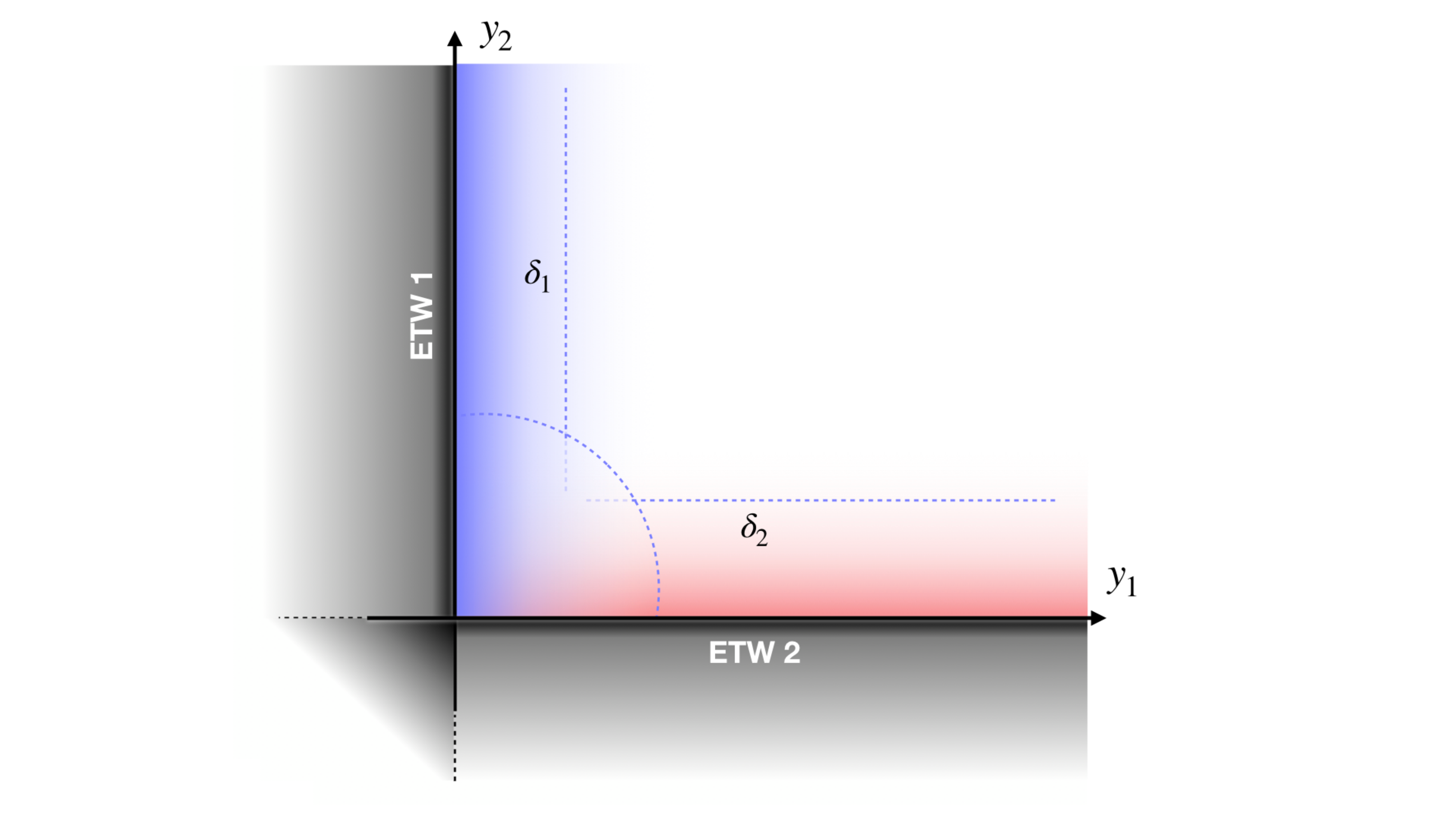}
\caption{\small Intersection of two orthogonal ETW branes of type $\delta_1$ and $\delta_2$. Our solutions zoom into the region near the intersection, denoted with a dashed curve.}
\label{fig:intersection}
\end{center}
\end{figure}

We are interested in solutions describing intersecting ETW branes, with the requirement that each scalar $\phi_i$ runs off to infinite distance in field space as it approaches the origin in the coordinate $y_i$ it depends on;  consequently both scalars diverge at the codimension-2 locus $y_1=y_2=0$. This is depicted in Figure \ref{fig:intersection}.

Let us emphasize again that the solution is intended to be a local description near the intersection point. Hence the center piece of Figure \ref{fig:intersection} is meant to describe a local patch near the intersection where this local description holds.

A simple proposal for the solution to describe two intersecting ETW branes is that for e.g. constant non-zero $y_2$, one recovers a codimension-1 ETW brane solution  (and similarly for $y_1$). Hence, the metric should be of the form (\ref{dw-ansatz}). This is achieved if we impose an additive structure in our warp factors
\begin{equation}
A(y_1,y_2)=-\sigma_1(y_1)-\sigma_2(y_2)\quad ,\quad B(y_1,y_2)=-\sigma_2(y_2)\quad ,\quad C(y_1,y_2)=-\sigma_1(y_1)  .  
\label{additive_str_warp}
\end{equation}
Note that we have not included additive factors depending on $y_1$ on $B$, or on $y_2$ on $C$, since they can be reabsorbed by redefining those coordinates. In fact, the above parametrization makes the connection with (\ref{dw-ansatz}) manifest. The metric (\ref{codim2-ansatz}) becomes
\begin{equation}
ds_{n+2}^2\,=\, e^{-2\sigma_1-2\sigma_2}ds_n^2 \,+ \, e^{-2\sigma_2}  dy_1^2\,+\, e^{-2\sigma_1}dy_2^2  \, ,
\label{restricted-ansatz-codim2}
\end{equation}
so that for e.g. in a constant non-zero $y_2$ slice, the resulting $(n+1)$-dimensional metric is (up to a constant)
\beqa
ds_{n+1}^2\,=\, e^{-2\sigma_1}ds_n^2 \,+ \, dy_1^2\,
\label{codim1-from-codim2}
\eeqa
with a running scalar $\phi_1(y_1)$, and $\phi_2$ remains constant along the slice. This is precisely the structure of the local description of a codimension-1 ETW brane. Obviously, a similar pattern holds for constant $y_1$ slices. Motivated by this, we can propose logarithmic profiles for the functions $\sigma_i$, $\phi_i$:
\beqa
&\sigma_1=-a_1 \log y_1 + \frac{1}{2} \log c_1, \quad \quad & \sigma_2=-a_2\log y_2 + \frac{1}{2} \log c_2 ,\nonumber \\
& \phi_1=b_1\log y_1, \quad \quad \quad \quad  \ \quad \quad & \phi_2=b_2\log y_2,
\label{log-ansatz}
\eeqa
where $c_1$ and $c_2$ correspond to (subleading) constant terms related to the two independent integration constants of the equations of motion.

Replacing these profiles in \eqref{lin_indep_eoms}, we get the following constraints:
\beqa
V&=&  - \frac{1}{2}c_1 n a_1 [\, (n+1) a_1 -1 \,] y_1^{-2} y_2^{-2a_2} - \frac{1}{2}c_2 n a_2 [\,(n+1)a_2 -1\, ] y_1^{-2a_1} y_2^{-2}\nonumber\\
&\equiv & -c_1v_1 y_1^{-2} y_2^{-2a_2}-c_2v_2 y_1^{-2a_1} y_2^{-2}\nonumber\\
 \alpha &=& 2\sqrt{a_1 a_2}\nonumber\\
b_i^2&=&na_i
\label{constraints-eoms}
\eeqa
Note the prefactors in the scalar potential, which are controlled by the constants $c_i$ in the logarithmic ansatz (\ref{log-ansatz}). In the first equation, assuming $a_1,a_2\neq 1$,  the potential splits in two pieces with different dependence on $y_1, y_2$. We thus split 
\beqa
V=V_1+V_2\quad, \quad 
V_1\sim -c_1v_1\,y_1^{-2} y_2^{-2a_2}\quad ,\quad V_2 \sim c_2v_2\, y_1^{-2a_1} y_2^{-2}
\eeqa
Note that the asymptotic behaviour of $V$ fixes the values of  $a_1,a_2$ and then there is no freedom to change $\alpha$. Hence this class of solutions requires a tuning of the mixed kinetic term for the scalars. 
This may seem a strong restriction on the theory; however, starting from a theory with e.g. no mixed terms, one can always redefine the scalars such that the appropriate mixed term arises. Hence the above condition can be regarded as specifying in which basis of the scalar fields the solution takes the above simple form. Interestingly we will show that the above set of solutions includes large classes of interesting examples, as we describe in section \ref{sec:examples}.

Using the above, we get
 \begin{equation}
  a_1= \frac{1 \pm \sqrt{1 +8v_1 \left( 1 + \frac{1}{n} \right)}}{2 (n+1)}   
\quad ,\quad
  a_2= \frac{1 \pm \sqrt{1 +8v_2 \left( 1 + \frac{1}{n} \right)}}{2 (n+1)}   
  \label{the-as}
 \end{equation}
In analogy with the codimension-1 ETW branes, we introduce the quantities $\delta_1$, $\delta_2$, such that the local ansatz reads
\beqa
V&=&V_1+V_2=c_1v_1 \, e^{\delta_1\phi_1}e^{a_2\delta_2\phi_2}
+c_2v_2\,e^{a_1\delta_1\phi_1}e^{\delta_2\phi_2}\nonumber\\
\phi_1&=&-\frac{2}{\delta_1}\log y_1\quad ,\quad \phi_2=-\frac{2}{\delta_2}\log y_2\nonumber \\
\sigma_1&=& -\frac{4}{n\delta_1^2} \log y_1\quad ,\quad \sigma_2=-\frac{4}{n\delta_2^2}\log y_2
\label{local-intersecting-etws}
\eeqa

This is the codimension-2 local description of intersecting ETW branes, which generalizes the structure of the local description for codimension-1 ETW branes in \cite{Angius:2022aeq}. The full solution is determined by the critical exponents $\delta_i$ associated with the two individual ETW branes. From the above equations, they are given by
 \begin{equation}
     \delta_1^2 = \frac{8 (n+1)}{n \pm \sqrt{n \left[ n+8v_1 (n+1) \right]}}
 \quad ,\quad
     \delta_2^2 = \frac{8 (n+1)}{n \pm \sqrt{n \left[ n+8v_2 (n+1) \right]}}.
     \label{critical-exp-intersecting-etws}
 \end{equation} 

Let us mention that our solution is even more general than what the derivation above suggests. Indeed, if one starts from (\ref{codim2-ansatz}) and requires a general additive structure with independent functions (i.e. beyond (\ref{additive_str_warp})) with general logarithmic profiles, the equations of motion end up leading to the above solution. Hence, (\ref{additive_str_warp}) can be regarded as a derived structure, once the logarithmic profiles are imposed.

We incidentally note that the ansatz (\ref{restricted-ansatz-codim2}) is conformally flat, generalizing the situation encountered in the codimension-1 case in section \ref{sec:codim1}. This is easily shown by changing to new coordinates $x^1$, $x^2$ such that
\beqa
dy_1=e^{-\sigma_1} dx_1\quad ,\quad dy_2=e^{-\sigma_2} dx_2
\eeqa
so that (\ref{restricted-ansatz-codim2}) becomes
\begin{equation}
ds_{n+2}^2\,=\, e^{-2\sigma_1-2\sigma_2}\,[\, ds_n^2 \,+ \,   dx_1^2\,+\,dx_2^2  \,]\, .
\label{conf-flat}
\end{equation}
where $\sigma_i$ are obviously regarded as functions of $x^i$. Conformally flat codimension-2 solutions have been discussed in setups with a single scalar e.g. in \cite{Dudas:2002dg}. It would be interesting to explore possible relations between the two setups. 

For our purposes, the expression (\ref{conf-flat}) will allow to generalize our solutions to ETW branes intersecting at general angles in appendix \ref{sec:angles}.
In coming sections we restrict to orthogonal intersections (\ref{restricted-ansatz-codim2})  for simplicity, the generalization being trivial.

\subsection{The sources}
\label{sec:sources}

In this section we discuss the source terms that correspond to our above intersecting ETW brane solutions, and check that they are just a superposition of source terms at $y_1=0$ and at $y_2=0$ closely related to the source terms of codimension-1 ETW branes. In particular, there is no further source term localized purely at the intersection $y_1=y_2=0$. For simplicity of the discussion, we focus on the case of zero scalar potential; the general case can be worked out similarly.

We start with recovering the source term for the codimension-1 ETW brane solutions following \cite{Blumenhagen:2023abk} (see also \cite{Blumenhagen:2000dc, Blumenhagen:2022mqw, Sugimoto:2023oul,Delgado:2023uqk}). We start with the $d$-dimensional action with a source term describing the tension and coupling to the scalar of the ETW brane:
\begin{equation}
\centering
    S = \int d^{d}x \sqrt{-g}\, \bigg[ \,\frac{1}{2} R - \frac{1}{2} \left( \partial \phi \right)^2\, \bigg] - \lambda \int d^{d-1}x\sqrt{-\tilde{g}}\,e^{\tilde{\alpha} \phi}\delta(y),
    \label{action_single_source}
\end{equation}
where now $\tilde{g}$ denotes the pullback of the $d$-dimensional metric $g$ to the worldvolume of the codimension-1 defect. Using the ansatz \eqref{dw-ansatz}, the two equations of motion coming from the variation with respect to the metric become:
\begin{equation}
\begin{split}
\{i,j\}: \ \ &\frac{1}{2}\phi'^2+\frac{(d-1)(d-2)}{2}\sigma'^2-(d-2)\sigma'' +\lambda e^{\tilde{\alpha}\phi}\delta(y)=0, \\
\{y,y\}: \ \ &\phi'^2-(d-1)(d-2)\sigma'^2=0,
\label{metric_var_y}
\end{split}    
\end{equation}
where the prime denotes derivation with respect to $y$. The variation of the action with respect to the field $\phi$ gives:
\begin{equation}
    \phi''-(d-1)\phi'\sigma'=\tilde{\alpha}\lambda e^{\tilde{\alpha}\phi}\delta(y).
\end{equation}
Making the ansatz $\phi = -\sqrt{\frac{d-2}{d-1}} f(y)$ and $\sigma = -\frac{1}{d-1} f(y)$ automatically satisfies the $\{y,y\}$ equation of motion, and the remaining two equations read:
\begin{equation}
\begin{split}
   &\frac{d-2}{d-1}\big(f'^2+f''\big)=-\lambda e^{\tilde{\alpha}\phi}\delta(y), \\
    &\sqrt{\frac{d-2}{d-1}}\big(f'^2+f''\big)=-\tilde{\alpha}\lambda e^{\tilde{\alpha}\phi}\delta(y).
    \end{split}
 \label{source-simple-codim1}
\end{equation}
By direct comparison one can read off the value of $\tilde \alpha$
\begin{equation}
    \tilde{\alpha}=\sqrt{\frac{d-1}{d-2}}.
    \label{scalar-coupling-codim1}
\end{equation}
Setting now $f(y)=\log(h(y))$ (with $h(y)\geq 0$) this becomes

\begin{equation}
\frac{h''}{h}=- \lambda \frac{d-1}{d-2}e^{-\log h}\delta(y)\, \quad \Rightarrow \quad \,h'' = - \lambda \, \frac{d-1}{d-2}\,\delta(y) \, .
\end{equation}

Now we can integrate this over $[0,x_0)$ and take the limit $x_0\to 0$. 
The left hand side is just the discontinuity of $h'$. Using the solution to the equations of motion, for $y<0$ we take $h=1$, so that $f(y)=0$ and all fields vanish beyond the ETW brane, while for $y>0$,  $h=y-y_0$, with $y_0$ an integration constant (with $y_0<0$ to have $h(y)\geq 0$ near $y=0$), which we will eventually take to $y_0\to 0^-$ to match our solution.

Then after the integration/limit we have:
\begin{equation}
    1=- \frac{d-1}{d-2} \lambda \quad \Rightarrow \quad \lambda = - \frac{d-2}{d-1}.
    \label{thelambda}
\end{equation}
The negative tension of this ETW brane was already explicitly noticed in \cite{Blumenhagen:2023abk} (see also \cite{Delgado:2023uqk}), for the physical choice of signs we have implicitly assumed in our solution.

We now turn to the intersecting ETW-brane solution, and show that codimension-1 sources of the kind studied above suffice to support the solution. We thus use a $d$-dimensional action (and set $d=n+2$) including sources of the following form:
\begin{equation}
\centering
\begin{split}
    S &= \int d^{n+2}x \sqrt{-g} \left\lbrace \frac{1}{2} R - \frac{1}{2} \left( \partial \phi_1 \right)^2 - \frac{1}{2} \left( \partial \phi_2 \right)^2 - \frac{\alpha}{2} \partial_{\rho} \phi_1 \partial^{\rho} \phi_2 \right\rbrace \\ &- \lambda_1 \int d^{n+1}x\sqrt{-g_1}e^{\alpha_{11} \phi_1+\alpha_{12}\phi_2}\delta(y_1)-\lambda_2 \int d^{n+1}x\sqrt{-g_2}e^{\alpha_{22} \phi_2+\alpha_{21}\phi_1}\delta(y_2)\, .
    \label{action_sources}
\end{split}
\end{equation}
where $g_i$, $i=1,2$ is the pullback of the metric on the worldvolume of the $(n+1)$-dimensional ETW branes.
Note that we have not included a term $\delta(y_1)\delta(y_2)$, as it is not necessary (in fact, it is forced to be absent) in our solution.

The equations of motion arising from the variation of the above action are very similar to those of \eqref{lin_indep_eoms}, when appropriately replacing the potential with the source terms. Using the ansatz \eqref{codim2-ansatz} for the metric and the additive structure \eqref{additive_str_warp} for the warp factors, the equations of motion coming from the variations of \eqref{action_sources} with respect to the metric components are:
\begin{equation}
\begin{split}
\{i,j\}: \ \ \frac{1}{2}\Big[ & e^{-2\sigma_1}\Big(\dot{\phi}_1^2+n(n+1)\dot{\sigma}_1^2-2n\ddot{\sigma}_1\Big)+e^{-2\sigma_2}\Big(\phi_2'^2+n(n+1)\sigma_2'^2-2n\sigma_2''\Big)\Big]\\ +&\lambda_1e^{\alpha_{11}\phi_1+\alpha_{12}\phi_2} e^{-2\sigma_1-\sigma_2}\delta(y_1) +\lambda_2e^{\alpha_{22}\phi_2+\alpha_{21}\phi_1} e^{-\sigma_1-2\sigma_2}\delta(y_2)   =0, \\  
\{y_1,y_1\}: \ \ &\frac{1}{2}\Big[ -\dot{\phi}_1^2+n(n+1)\dot{\sigma}_1^2+e^{2\sigma_1-2\sigma_2}\big(\phi_2'^2+n(n+1)\sigma_2'^2-2n\sigma_2''\big)\Big]\\ & \ \ \ \ \ +\lambda_2e^{\alpha_{22}\phi_2+\alpha_{21}\phi_1} e^{-2\sigma_2+\sigma_1}\delta(y_2)=0, \\  
\{y_2,y_2\}: \ \ &\frac{1}{2}\Big[ -\phi_2'^2+n(n+1)\sigma_2'^2+e^{-2\sigma_1+2\sigma_2}\big(\dot{\phi}_1^2+n(n+1)\dot{\sigma}_1^2-2n\ddot{\sigma}_1\big)\Big]\\ & \ \ \ \ \ +\lambda_1e^{\alpha_{11}\phi_1+\alpha_{12}\phi_2} e^{-2\sigma_1+\sigma_2}\delta(y_1)=0, \\  
\{y_1,y_2\}:\ \ & -n\dot{\sigma}_1\sigma_2'+\frac{1}{2}\alpha\dot{\phi}_1\phi_2'=0.
\end{split} 
\label{sources1}
\end{equation}

Additionally, one has two equations of motion coming from the variations with respect to the fields $\phi_i$:
\begin{equation}
\begin{split}
\phi_1: \ \ & e^{-\sigma_1+\sigma_2} \big(-(n+1)\dot{\phi}_1\dot{\sigma}_1+\ddot{\phi}_1\big)+\frac{\alpha}{2} e^{\sigma_1-\sigma_2} \big(-(n+1)\phi_2'\sigma_2'+\phi_2''\big)\\-& \alpha_{11} \lambda_1 e^{\alpha_{11}\phi_1+\alpha_{12}\phi_2} e^{-\sigma_1}\delta(y_1) -
\alpha_{21} \lambda_2 e^{\alpha_{22}\phi_2+\alpha_{21}\phi_1} e^{-\sigma_2}\delta(y_2)
=0, \\  
\phi_2: \ \ &\frac{\alpha}{2} e^{-\sigma_1+\sigma_2} \big(-(n+1)\dot{\phi}_1\dot{\sigma}_1+\ddot{\phi}_1\big)+ e^{\sigma_1-\sigma_2} \big(-(n+1)\phi_2'\sigma_2'+\phi_2''\big)\\-& \alpha_{12} \lambda_1 e^{\alpha_{11}\phi_1+\alpha_{12}\phi_2} e^{-\sigma_1}\delta(y_1) -
\alpha_{22} \lambda_2 e^{\alpha_{22}\phi_2+\alpha_{21}\phi_1} e^{-\sigma_2}\delta(y_2)
=0 .   
\end{split}  
\label{sources2}
\end{equation}

In analogy with the codimension-1 case, let us make the change $\phi_i=-\sqrt{\frac{n}{n+1}} f_i(y_i)$, $\sigma_i=-\frac{1}{n+1}f_i$ 
to simplify the equations, 

in particular the $\{y_1,y_2\}$ equation in (\ref{sources1}) is satisfied identically. Each of the remaining equations splits into two contributions depending on each the two coordinates, provided that $\alpha_{21}\phi_1=\sigma_1$ and  $\alpha_{12}\phi_2=\sigma_2$,
hence
\beqa
\alpha_{12}=\alpha_{21}=\frac{1}{\sqrt{n(n+1)}} .
\eeqa

For the scalar equations (\ref{sources2}) to be compatible one needs
\beqa
\alpha = \frac 2{n+1} .
\eeqa
which corresponds to the appropriate value for the case of vanishing potential. The set of equations decouples into two sets basically identical to the codimension-1 equations (\ref{source-simple-codim1}) for the $f_i$, namely
\begin{equation}
\begin{split}
   & \frac{n}{n+1} \left( f_i'^2+f_i'' \right) =-\lambda_i e^{\alpha_{ii} \phi_i} \delta(y_i), \\
    & \sqrt{\frac{n}{n+1}} \left( f_i'^2+f_i'' \right) =-\alpha_{ii} \lambda_i e^{\alpha_{ii} \phi_i} \delta(y_i)  ,
\end{split}
 \label{source-simple-codim2}
\end{equation}

\noindent
with no sum over $i$, and where prime denotes derivative with respect to their argument, now also for $f_1(y_1)$. The equations can be analyzed as in the codimension-1 case, so their compatibility requires
\beqa
\alpha_{11}=\alpha_{22}=\sqrt{\frac{n+1}{n}}=\sqrt{\frac{d-1}{d-2}}
\eeqa
and the computation of the discontinuities imply
\beqa
\lambda_1=\lambda_2=-\frac{n-1}{n}=-\frac{d-2}{d-1} , 
\eeqa 
just like in the codimension-1 case, c.f. (\ref{thelambda}). 

Hence the sources are a simple superposition of two terms along $y_i=0$, with tensions $\lambda_i$ and couplings $\alpha_{ii}$ to the scalar $\phi_i$ given by those of the corresponding codimension-1 ETW brane solution. The extra coupling $\alpha_{12}$ of the ETW brane along $y_1=0$ to $\phi_2$, and $\alpha_{21}$ of the ETW along $y_2=0$ to $\phi_1$ imply an interesting variation of the effective tension of the ETW branes as one moves further away from the intersection. Morally, it accounts for the extra factor involved in expressing the codimension-2 solution as a codimension-1 solution, mentioned just above (\ref{codim1-from-codim2}).

Let us finally explain the solution leaves no room for a codimension-2 $\delta(y_1)\delta(y_2)$ source term. Such term would lead to discontinuities in mixed derivatives of the fields, which are absent in the equations of motion (\ref{sources1}), (\ref{sources2}). This is already built in from the use of the additive structure (\ref{additive_str_warp}). It would be interesting to explore more general solutions involving this extra sources, but this lies beyond the scope of this work.

\subsection{Scaling relations}
\label{sec:scalings}

In this section we discuss the analogue of the scaling relations (\ref{universal-scalings-codim1}), in particular the relation between the spacetime distance to the singularity along some path and the traversed scalar field space distance. Clearly, the relation will be path-dependent, albeit in a simple way.

Consider a general path $y_i(t)$ in spacetime, parametrized by $t$, with $y^i\to 0$ as $t\to 0$. 
For instance, we can choose
\beqa
y_1=t^{\gamma_1}\quad , \quad y_2=t^{\gamma_2}
\label{path-parameterization}
\eeqa
in terms of two positive real numbers $\gamma_i\geq 0$. Clearly, a change $t\to t^\lambda$ is just a reparametrization of the same path, so $\gamma_i$ are defined up to an overall rescaling, so only its ratio is meaningful. The tangent vector is
\beqa
\partial_t y_i= \gamma_i \,t^{\gamma_i-1} .
\label{tangent-vector}
\eeqa
The spacetime distance to the origin along this path is
\beqa
\Delta&=&\int[\, e^{-2\sigma_2} (\partial_t y_1)^2 + e^{-2\sigma_1}(\partial_t y_2)^2
\,]^{\frac 12}\,dt\, =\, 
\, \int[\, \gamma_1^2 t^{2r_1} + \gamma_2^2 t^{2r_2}\,]^{\frac 12}\, dt
\label{distance-element}
\eeqa
with
\beqa
r_1=\frac{4\gamma_2}{n\delta_2^2}+\gamma_1-1\quad ,\quad 
r_2=\frac{4\gamma_1}{n\delta_1^2}+\gamma_2-1
\label{the-rs}
\eeqa
We can consider two regimes, depending on which contribution dominates in the $t\to 0$ limit. The two regions are separated by the line $r_1=r_2$, equivalently
\beqa
\frac{\gamma_1}{\gamma_2}\,=\,\frac{\frac{4}{n\delta_2^2}-1}{\frac{4}{n\delta_1^2}-1} .
\label{gamma-ratio}
\eeqa
The two regimes correspond to the path being closer to each of the two ETW branes. Assuming $r_1\neq r_2$, the spacetime distance in the two regions is given by
\beqa
\Delta= \int \gamma_i \,t^{r_i}\,dt=\frac{\gamma_i}{r_i+1} t^{r_i+1}.
\label{spacetime-distance-tworegions}
\eeqa

We incidentally note that the $r_i$ have a natural interpretation by writing the tangent vector (\ref{tangent-vector}) in the tangent space frame\footnote{We warn the reader that we are using lowercase indices for our spacetime coordinates $y_i$, which leads to some funny contraction of indices, like in this formula.} $\tau^a=e_i^a\partial_t y_i$ with $e_i^a$ defined by $G_{ij}=e_i^ae_j^b \delta_{ab}$
\beqa
\vec{\tau}=(e^{-\sigma_2} \partial_t y_1, e^{-\sigma_1}\partial_t y_2)=(\gamma_1 t^{r_1}, \gamma_2 t^{r_2}) .
\eeqa
Products of vectors in the tangent space are with the flat metric $\delta_{ab}$, so this reproduces the distance element (\ref{distance-element}). The tangent vector $\vec{\tau}$ will play an interesting role in the discussion of the Distance Conjecture in section \ref{sec:swamp-distance}.

The profiles for the scalars allow to translate the path $y_i(t)$ in spacetime into a path in field space $\phi_i(y(t))$.
Let us now compute the field theory distance traversed along the path from a point located at some small non-zero value $t$ to the origin $t=0$. From the action (\ref{action_2scalars}), the line element in field space is given by
\beqa
d{\cal D}^2&=&d\phi_1^2+d\phi_2^2+\alpha d\phi_1 d\phi_2=
4\left(\frac{\gamma_1^2}{\delta_1^2}+\frac {\gamma_2^2}{\delta_2^2}+\frac{\alpha \gamma_1 \gamma_2}{\delta_1\delta_2}\right)\frac{dt^2}{t^2} .
\eeqa
This can be recast in terms of the spacetime distance, for each of the two regions in (\ref{spacetime-distance-tworegions}), as
\beqa
{\cal D}= -2\left(\frac{\gamma_1^2}{\delta_1^2}+\frac {\gamma_2^2}{\delta_2^2}+\frac{\alpha \gamma_1 \gamma_2}{\delta_1\delta_2}\right)^{\frac 12} \log t  \sim  - \frac 2{|r_i+1|}\left(\frac{\gamma_1^2}{\delta_1^2}+\frac {\gamma_2^2}{\delta_2^2}+\frac{\alpha \gamma_1 \gamma_2}{\delta_1\delta_2}\right)^{\frac 12}\log \Delta . \quad \quad  
\label{moduli-distance}
\eeqa
The explicit dependence on the parametrization i.e. on the $\gamma_i$, is simply due to the fact that the choice of initial point for the computation of the distance in terms of a value $t$ does depend on the parametrization (\ref{path-parameterization}).

We thus obtain a scaling relation near the intersection, of the kind (\ref{universal-scalings-codim1}), namely
\beqa
\Delta\sim e^{-\frac 12 \delta_{\rm int}\, {\cal D}}\, ,
\label{scaling-codim2}
\eeqa
with a path-dependent coefficient, which in each of the two regions reads
\beqa
\delta_{\rm int}= \left(\frac{\gamma_1^2}{\delta_1^2}+\frac {\gamma_2^2}{\delta_2^2}+\frac{\alpha \gamma_1 \gamma_2}{\delta_1\delta_2}\right)^{-\frac 12}(r_i+1).
\label{delta-int}
\eeqa
This interpolates between the values of the critical exponents of the two individual ETW branes, which are attained for paths orthogonal to the individual ETW branes:
\begin{equation}
\begin{split}
    \gamma_1 = 1 \; , &  \; \gamma_2 =0 \quad \quad \Rightarrow \quad \delta_{\rm int}\to \delta_1 \\
    \gamma_1 = 0 \; , & \; \gamma_2 =1 \quad \quad \Rightarrow \quad \delta_{\rm int}\to \delta_2  \\
\end{split}    
\end{equation}

In order to visualize the interpolation, let us parametrize $\gamma_1= \sqrt{\gamma}$ and $\gamma_2= 1/ \sqrt{\gamma}$ such that we have the ratio $\gamma_1/\gamma_2=\gamma$. The value of $\gamma$ separating the two regimes is (\ref{gamma-ratio}).
In terms of this parametrization, we have
\begin{equation}
    \delta_{int} = \begin{cases}
        & \left( \frac{\gamma}{\delta_1^2} + \frac{1}{\gamma  \delta_2^2} + \frac{\alpha}{\delta_1 \delta_2} \right)^{-1/2} \left( \sqrt{\gamma} + \frac{4}{n \delta_2^2 \sqrt{\gamma}}\right) \quad \quad \quad \text{for } \gamma>\gamma^{\ast} \\
         & \left( \frac{\gamma}{\delta_1^2} + \frac{1}{\gamma \delta_2^2} + \frac{\alpha}{\delta_1 \delta_2} \right)^{-1/2} \left( \frac{1}{\sqrt{\gamma}} + \frac{4 \sqrt{\gamma}}{n \delta_1^2 }\right) \quad \quad \quad  \quad \text{for } \gamma<\gamma^{\ast}. \\
    \end{cases}
\end{equation}
Note that we recover the limits $\delta_1$, $\delta_2$ for  $\gamma \mapsto \infty, 0$, respectively.

\begin{figure}
    \centering
    \includegraphics[scale=0.3]{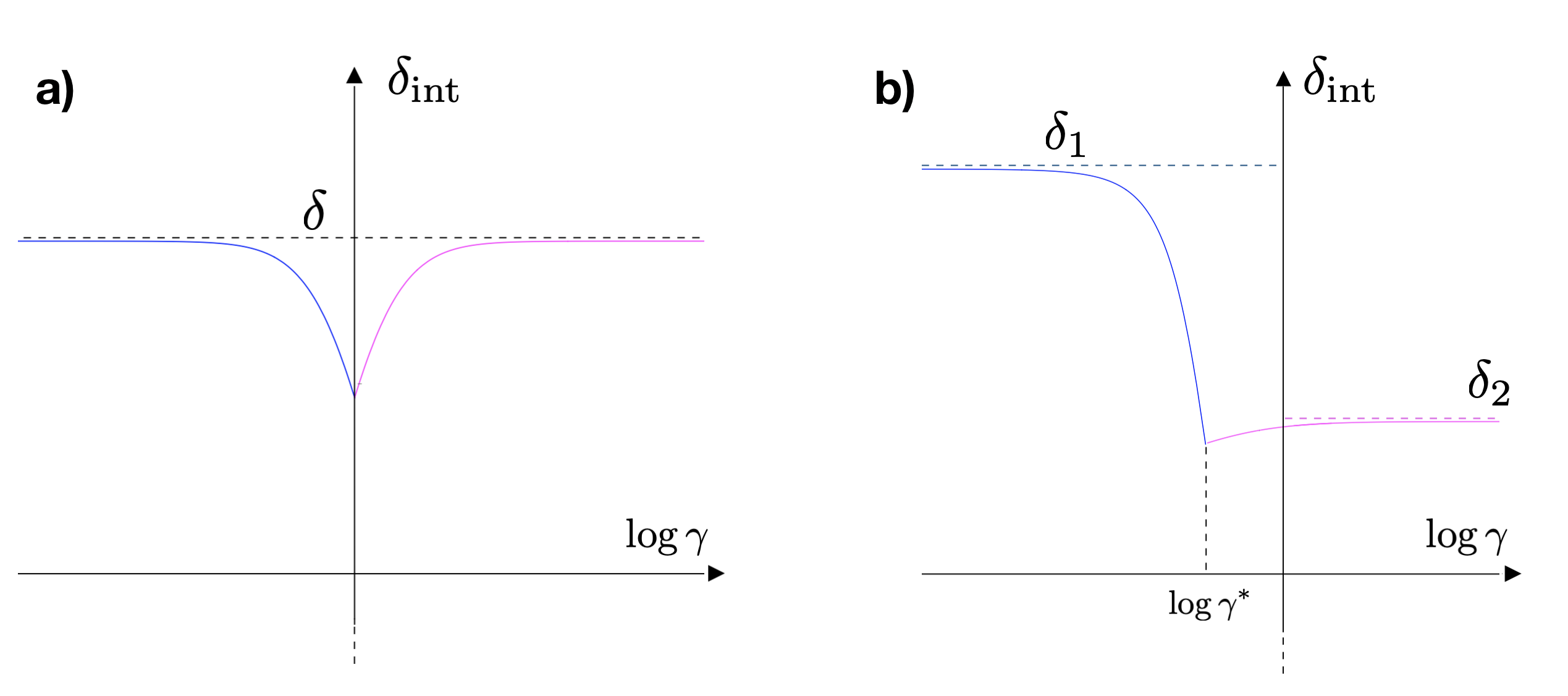}
    \caption{\small Plot of $\delta_{int}$ as a function of the path for two illustrative examples with $\delta_1= \delta_2\equiv\delta$} (figure a) and $\delta_1\neq\delta_2$ (figure b).
    \label{fig:plots}
\end{figure}

Consider for instance the case $\delta_1=\delta_2\equiv \delta$. The two regimes are separated by $\gamma^{\ast}=1$ and the scaling parameter simplifies:
\begin{equation}
    \delta_{int}= \begin{cases}
        & \left[ 1 + \gamma(\gamma+ \alpha) \right]^{-1/2} \left( \frac{4+n \gamma \delta^2}{n \delta} \right) \quad \quad \text{for } \gamma>1 \\
       & \left[ 1 + \gamma(\gamma+ \alpha) \right]^{-1/2} \left( \frac{4\gamma+n \delta^2}{n \delta} \right) \quad \quad \text{for } \gamma<1. \\  
    \end{cases}
\end{equation}
For $\delta_1\neq \delta_2$ the separation between the two regimes lies at $\gamma^\ast\neq 1$. In Figures \ref{fig:plots} we display $\delta_{\rm int}$ as a function of $\log\gamma$, for some illustrative examples with equal (figure a) or different (figure b) values of $\delta_1,\delta_2$.

The scaling properties between the spacetime and field theory distance will play an important role in the discussion of swampland conjectures in section \ref{sec:swampland}. In the following section we turn to show several explicit examples of systems described by the solution we have discussed. 

\section{Explicit examples}
\label{sec:examples}

In this section we consider explicit examples of intersecting ETW brane solutions. Many are simply obtained from simple configurations, like flat space, by considering reduction along some isometry orbits, which potentially diverge at large distances from the intersection. The solutions we describe should be regarded as local descriptions near the intersection of more involved solutions where such orbits have finite size at infinity. We will mention explicit realizations for several of our examples.

\subsection{${\bf S}^1\times {\bf S}^1$ compactifications}
\label{example-s1s1}

A simple example of a codimension-1 ETW brane is the wall of nothing in $\IS^1$ compactifications (analogue of the bubble of nothing in \cite{Witten:1981gj}) described in section \ref{sec:codim1}. From the higher-dimensional perspective, the local description corresponds to taking flat space, splitting of an $\IR^2$, and slicing it along the angular $\IS^1$, regarding it as a compactification circle, with radius varying along the radial coordinate, c.f. (\ref{flat-s1-sliced}).

In the same spirit, we can consider the intersection of two wall of nothing ETW branes, which provides a local model for two\footnote{The case of two intersecting bubbles of nothing for the same $\IS^1$ belongs to the class of solutions considered in section \ref{sec:single}.} intersecting bubbles of nothing for different $\IS^1$'s. The idea is simply to consider the flat $(n+4)$-dimensional space, written as
\beqa
ds_{n+4}^2=ds_n^2+dr_1^2+dr_2^2+r_1^2d\theta_1^2+r_2^2 d\theta_2^2
\label{flat-s1s1-slicing}
\eeqa
where $ds_n^2$ is just flat $n$-dimensional space.

The above will soon be rewritten in the Einstein frame of the $(n+2)$-dimensional theory obtained upon reduction on the $\IS^1\times \IS^1$ parametrized by $\theta_i$; but the intuition is already clear. We have an $(n+2)$-dimensional theory with two scalars, the $\IS^1$ sizes, depending of two coordinates $r_i$. Each scalar shrinks to zero size at the codimension-1 locus $r_1=0$ or $r_2=0$, respectively, and both shrink simultaneously at the codimension-2 locus $r_1=r_2=0$.

We now turn to carrying this out explicitly. 
We start with $(n+4)$-dimensional gravity with action
\beqa
 S_{n+4} = \frac{1}{2} \int d^{n+4}x \sqrt{-g_{n+4}} \,R_{n+4}
\eeqa
and compactify on $\IS^1\times\IS^1$, parametrized by coordinates $\theta_1$, $\theta_2$. We consider the following ansatz for the $(n+2)$-dimensional theory:
\beqa
ds^2_{n+4} = e^{\alpha_1\rho_2+\alpha_2\rho_2} ds^2_{n+2} +e^{-\beta_1\rho_1} d \theta_1^2 + e^{-\beta_2\rho_2} d \theta_2^2,
    \label{compact_ansatz1}
\eeqa
where the breathing modes $\rho_i$ are functions of the non-compact $n+2$ dimensions\footnote{Although in this particular example they will turn out to be the relevant lower dimensional scalars $\phi_i$, we choose to maintain a specific notation for breathing modes, as in general there may be additional components entering the scalars $\phi_i$ (see sections \ref{sec:dbrane}, \ref{sec:one-general-ETW} for examples).}. The parameters are fixed by the $(n+2)$-dimensional Einstein frame condition, and normalization of the scalar kinetic terms, giving
\beqa
 n \alpha_i =  \beta_i   \quad , \quad \beta_i^2=\frac{4n}{n+1} .
\eeqa 
Upon compactification, the $(n+2)$-dimensional action for these fields is
\beqa
         S_{n+2}& \propto &\frac{1}{2} \int d^{n+2}x \sqrt{-g_{n+2}} \big[\, R_{n+2} - \vert d \rho_1 \vert^2 - \vert d \rho_2 \vert^2 -\alpha \partial_{\mu} \rho_1 \partial^{\mu} \rho_2\,\big]
        \, ,
    \label{action_S1xS1_radions}
\eeqa
with 
\beqa
\alpha= \frac{2}{n+1}.
\eeqa

This corresponds to an action of the kind (\ref{action_2scalars}), with the two scalars corresponding to the $\IS^1$ sizes, namely $\phi_i\equiv \rho_i$. The scalar potential is zero because the $\IS^1$'s have no curvature. 

The flat space slicing (\ref{flat-s1s1-slicing}) corresponds to an intersecting ETW brane solution of the kind in section \ref{sec:intersecting} with
\beqa
 ds^2_{n+2} &=& r_1^{\frac{2}{n}} r_2^{\frac{2}{n}} \left( \eta_{\mu \nu} dx^{\mu} dx^{\nu} + dr_1^2 + dr^2_2\right), \nonumber\\
\phi_1&=&- \sqrt{\frac{n+1}{n}}\log r_1 \quad,\quad \phi_2= - \sqrt{\frac{n+1}{n}} \log r_2 .
\eeqa
With a change of variables 
\beqa
y_i=\,\frac{n}{n+1}\,r_i^{\frac{n+1}n}\, ,
\eeqa
we can go from the above conformally flat solution to one of the form (\ref{codim2-ansatz}) with
\beqa
ds_{n+2}^2 &= &y_1^{\frac{2}{n+1}}y_2^{\frac{2}{n+1}}\, ds_n^2\,+\,y_2^{\frac{2}{n+1}}dy_1^2\,+\, y_1^{\frac{2}{n+1}}dy_2^2\nonumber , \\
\phi_i&=& -\sqrt{\frac{n}{n+1}}\log y_i .
\eeqa
This corresponds to the critical exponents
\beqa
\delta_i=2\sqrt{\frac{n+1}{n}} .
\eeqa

\subsection{$\IS^{p_1}\times\IS^{p_2}$ compactifications}
\label{sec:sp1sp2}

We now consider a generalization of the above example, by starting with an $D$-dimensional space, with $D=n+2+p_1+p_2$, and compactifying on $\IS^{p_1}\times\IS^{p_2}$. The intersecting ETW brane solution, in which the $\IS^{p_1}$ and $\IS^{p_2}$ shrink to zero size at two interescting codimension-1 loci, is locally given by slicing $D$-dimensional flat space as
\beqa
 ds^2_D = \eta_{\mu\nu} dx^\mu dx^\nu + dr_1^2 +dr_2^2 + r_1^2 d \Omega_{p_1}^2 + r_2^2 d \Omega_{p_2}^2,
 \label{flat-spsq-slicing}
\eeqa
where $x^\mu$ are coordinates along the Poincar\'e invariant directions along the intersection and $d\Omega_{p_1}^2$ and $d \Omega_{p_2}^2$ are the line elements in $\IS^{p_1}$ and $\IS^{p_2}$, respectively.

Let us quickly describe this construction. We start with $D$-dimensional gravity with action
\beqa
S_D = \frac{1}{2} \int d^D x \sqrt{-g_D} R_D.
\eeqa
The compactification ansatz is
\beqa
ds^2_D = e^{\alpha_1\rho_2+\alpha_2\rho_2} ds^2_{n+2} +e^{-\beta_1\rho_1} d \Omega^2_{p_1} + e^{-\beta_2\rho_2} d \Omega^2_{p_2},
    \label{ansatz_SpxSq}
\eeqa
with the requirement to land on the $(n+2)$-dimensional Einstein frame leading to
\beqa
 n \alpha_i =  p_i \beta_i   .
 \eeqa
Also, we normalize the kinetic terms of the two radions $\rho_i$ via the following relations:
\beqa
\beta_i^2   =\frac{4n}{p_i(n+p_i)} 
\eeqa 
The $(n+2)$-dimensional Einstein frame action for gravity and the two radions is
\beqa
         S_{n+2}& \propto &\frac{1}{2} \int d^{n+2}x \sqrt{-g_{n+2}} \Big[\, R_{n+2} - \vert d \rho_1 \vert^2 - \vert d \rho_2 \vert^2 -\alpha \partial_{\mu} \rho_1 \partial^{\mu} \rho_2\nonumber\\
       && +\frac{p_1(p_1-1)}{R_{p_1}^2} e^{(\alpha_1 + \beta_1) \rho_1 +\alpha_2  \rho_2} +\frac{p_2(p_2-1)}{R^2_{p_2}} e^{\alpha_1 \rho_1 +(\alpha_2+ \beta_2) \rho_2} \Big] \, ,
    \label{action_SpxSq_radions}
\eeqa
with
\beqa
\alpha= \frac{2{\sqrt{p_1p_2}}}{\sqrt{(n+p_1)(n+p_2)}}.
\eeqa
The action (\ref{action_SpxSq_radions}) has precisely the structure (\ref{action_2scalars}) with the two radions corresponding to the two scalars i.e. $\phi_i\equiv \rho_i$, which have an exponential potential due to the curvature of the internal spheres.

The flat space slicing (\ref{flat-spsq-slicing}) provides a solution to this theory with the structure (\ref{restricted-ansatz-codim2}), (\ref{log-ansatz}) with
\beqa
ds_{n+2}^2 &= &y_1^{\frac{2p_1}{p_1+n}}y_2^{\frac{2p_2}{p_2+n}}\, , ds_n^2\,+\,y_2^{\frac{2p_2}{p_2+n}}dy_1^2\,+\, y_1^{\frac{2p_1}{p_1+n}}dy_2^2 , \nonumber \\
\phi_i&=& -\sqrt{\frac{np_i}{n+p_i}}\log y_i .
\eeqa
This corresponds to the critical exponents
\beqa
\delta_i=2\sqrt{\frac{n+p_i}{np_i}} .
\label{delta-sp1sp2}
\eeqa
The $\IS^1\times\IS^1$ example in section \ref{example-s1s1} is clearly recovered for $p_1=p_2=1$.

One particular example realizing this local behaviour (albeit with AdS$_4$ rather than 4d Poincar\'e invariance along the ETW brane) is the gravity dual of 4d $\NN=4$ $SU(N)$ SYM with a  boundary coupled to a 3d $\NN=4$ BCFT \cite{Gaiotto:2008sa,Gaiotto:2008ak}. It is given by the supergravity solution of D3-branes ending on NS5- and D5-branes), see \cite{DHoker:2007zhm,DHoker:2007hhe,Aharony:2011yc,Assel:2011xz,Assel:2012cj,Bachas:2017rch,Bachas:2018zmb,Raamsdonk:2020tin,VanRaamsdonk:2021duo,Huertas:2023syg}, which as emphasized in \cite{Raamsdonk:2020tin,VanRaamsdonk:2021duo,Huertas:2023syg} corresponds to an ETW brane in AdS$_5\times\IS^5$. It is described as an AdS$_4$ times $(\IS^2)^2$ fibered over a Riemann surface given by the first quadrant $(y_1,y_2)$ with $y_i\geq 0$. The two $\IS^2$'s shrink to zero size at $y_1=0$ and $y_2=0$ respectively, while both shrink simultaneously at $y_1=y_2=0$ (so, together with the polar angle in the $(y_1,y_2)$-plane, there is a shrinking $\IS^5$). Hence, this setup reproduces locally the structure of our solution, while globally provides an explicit example in which the $\IS^2$'s have constant radius at infinity (being part of the constant radius $\IS^5$ in the asymptotic AdS$_5\times\IS^5$).

\subsection{Adding D-brane defects}
\label{sec:dbrane}

The above examples correspond to variants of bubbles of nothing, in which the ETW brane is realized geometrically, by shrinking (parts of) the compact space. In general, one expects ETW branes to be dressed with topological defects necessary to remove non-trivial cobordism charges of the compactification. A prototypical example is compactifications with field strength fluxes, which require the presence of charged branes at the ETW brane to remove the flux. In this section we consider a simple example describing the local behaviour of an ETW brane for a compactification on $\IS^{8-p}$  with $N$ units of RR flux (hence the ETW brane is dressed with $N$ D$p$-branes) intersecting an ETW brane of a fluxless $\IS^q$ (i.e. a bubble of nothing).

Actually, the solution is described by simply taking the solution for a stack of $N$ D$p$-branes in flat space. For simplicity we focus on $p<7$, for which the metric and dilaton read
\begin{equation}
    \begin{split}
        &ds^2_{10} = Z (r_1)^{\frac{p-7}{8}} (\eta_{\mu \nu} dx^{\mu} dx^{\nu}\,+ \, dr_2^2 +r_2^2d\Omega_q^2) + Z (r_1)^{\frac{p+1}{8}} \left( dr_1^2 + r_1^2 d \Omega^2_{8-p} \right), \\
        & \Phi = \frac{(3-p)}{4 \sqrt{2}} \log Z(r_1),\\
        & Z(r_1) = 1 + \left( \frac{\rho}{r_1} \right)^{7-p} \quad ,\quad \rho^{7-p}=g_s N \alpha'^{(7-p)/2} (4 \pi)^{(5-p)/2} \Gamma \left( \frac{7-p}{2} \right)\, .
    \end{split}
    \label{Dpbrane_solution}
\end{equation}
Here $x^\mu$ parametrize $p-q$ of the D$p$-brane worlvolume directions, and $r_2$ is the radial coordinate in the remaining $\IR^{p+1}$ part of the worldvolume, and $d\Omega_q^2$ is the line element in the angular $\IS^q$.
Namely, we slice the D$p$-brane solution along the transverse $\IS^{8-p}$ and a worldvolume $\IS^q$, and regard it as a solution of an $\IS^{8-p}\times\IS^q$ compactification.  

We thus start with the  10d action for gravity coupled to the dilaton and the RR $(p+1)$-form
\begin{equation}
    S_{10} \sim \frac{1}{2} \int d^{10}x \sqrt{-g_{10}} \left\lbrace R_{10} - \vert d \Phi \vert^2 - \frac{1}{2(8-p)! } e^{a \Phi} \vert F_{8-p} \vert^2 \right\rbrace
    \label{action_Dp-brane}
\end{equation}
with
\begin{equation}
    a= \frac{p-3}{2}.
\end{equation}
We use the compactification ansatz
\begin{equation}
    \begin{split}
        & ds^2_{10} = e^{\alpha_1 \rho_1 + \alpha_2 \rho_2} ds^2_{n+2} + e^{- \beta_1 \rho_1} d \Omega^2_{8-p} + e^{\gamma_1 \rho_1 - \gamma_2 \rho_2} d \Omega^2_q \\
        & F_{8-p} = {N} dVol \left( S^{8-p} \right).
    \end{split}
    \label{ansatz_compactification_Dpbrane}
\end{equation}
We now impose the Einstein frame condition, and fix the normalization of the scalars $\rho_i$
\begin{equation}
   \begin{split}
       &\beta_1 = \frac{n \alpha_1 +q \gamma_1}{8-p} = \frac{1}{8+n-p} \left[ (p-n) \gamma_1 \pm 2 \frac{\sqrt{n\left[ 8+n-p+2(p-n) \gamma_1^2 \right]}}{\sqrt{8-p}} \right]\\
       & \gamma_{2}=\frac{n}{q} \alpha_2 = \pm 2 \sqrt{ \frac{n}{q(n+q)}}\\
   \end{split}
\end{equation}
The resulting $(n+2)$-dimensional action (with $n=p-q$) is:
\begin{equation}
 \begin{split}
     & S_{n+2} =   \frac{1}{2}\, {\cal C}\! \int  d^{n+2} x \sqrt{-g_{n+2}} \left\lbrace R_{n+2}  - \vert d \Phi \vert^2 - \vert d \rho_1 \vert^2 - \vert d \rho_2 \vert^2 + \frac{\gamma_2 q \left[ \beta_1 (p-8)+\gamma_1 p \right]}{2 n} \partial_{\mu} \rho_1 \partial^{\mu} \rho_2 + \right. \\
     & \left. + (8-p)(7-p) e^{(\alpha_1 + \beta_1) \rho_1 +\alpha_2  \rho_2}  +q(q-1) e^{(\alpha_1-\gamma_1) \rho_1 +(\alpha_2 + \gamma_2) \rho_2}  - \frac{ N^2 }{2(8-p)!} e^{a \Phi} e^{ \left[ \alpha_1 +(8-p) \beta_1 \right] \rho_1+  \alpha_2 \rho_2}  \right\rbrace, \\
\end{split}   
\end{equation}

with the coefficient
\beqa
{\cal C}=\left( \frac{2 \pi^{(8-p)/2}}{\Gamma \left( \frac{(8-p)}{2} \right)} \right) \left( \frac{2 \pi^{q/2}}{\Gamma \left( \frac{q}{2} \right)} \right).
\eeqa

\noindent
We can now write this in the form (\ref{action_2scalars}) using the redefinitions
\beqa
\phi_1&=& \int \left[ \vert d \Phi \vert^2 + \vert d \rho_1 \vert^2 \right]^{1/2}\nonumber\\
\phi_2&=& \rho_2
\eeqa
and the following relation between the $10$ dimensional dilaton and the radion:
\begin{equation}
    \Phi = \sqrt{2} \frac{(p-7)}{(p-3)} \rho_1,
\end{equation}
from which we have:
\begin{equation}
    \phi_1 = \left[1 -2 \frac{\gamma_1}{\beta_1}\right]^{1/2} \rho_1 .
\end{equation}
The action becomes:
\begin{equation}
    \begin{split}
        & S_{n+2} = \frac{\mathcal{C}}{2} \int d^{n+2}x \sqrt{-g_{n+2}} \left\lbrace R_{n+2} - \vert d \phi_1 \vert^2 - \vert d \phi_2 \vert^2 -\alpha \partial_{\rho} \phi_1 \partial^{\rho} \phi_2 + \right. \\
        & \left. + \frac{(8-p)(7-p)}{R^2_{8-p}} e^{\frac{1}{n} \left( 1 -2 \frac{\gamma_1}{\beta_1} \right)^{-1/2} \left[ (8+n-p) \beta_1 +(n-p) \gamma_1 \right] \phi_1} e^{2 \sqrt{\frac{q}{n(n+q)}} \phi_2} \right. + \\
        & +\frac{q(q-1)}{R^2_q} e^{\frac{1}{n} \left( 1 -2 \frac{\gamma_1}{\beta_1} \right)^{-1/2} \left[(8-p) \beta_1 -p \gamma_1 \right] \phi_1} e^{2 \sqrt{\frac{n+q}{nq}} \phi_2} +\\
        & \left. - \frac{N^2}{2(8-p)!} e^{ \left\lbrace \frac{(p-3)}{2} \left(1 - \frac{1}{2} \frac{\beta_1}{\gamma_1} \right)^{-1/2}- \frac{1}{n} \left( 1-2 \frac{\gamma_1}{\beta_1} \right)^{-1/2} \left[ (n+1)(p-8) \beta_1 +q \gamma_1 \right] \right\rbrace \phi_1} e^{2 \sqrt{\frac{q}{n(n+q)}} \phi_2} \right\rbrace\\
    \end{split}
\end{equation}
with
\begin{equation}
    \alpha= \frac{\gamma_2 q \left[ \beta_1 (8-p)- \gamma_1 p \right]}{2n} \left( 1-2 \frac{\gamma_1}{\beta_1} \right)^{-1/2}.
\end{equation}
The D$p$-brane solution descends to a solution of the kind (\ref{codim2-ansatz}) with the redefinitions
\begin{equation}
\begin{split}
& y_1= \left( \frac{2n}{9+n(p-5)-p} \right) r_1^{\frac{9+n(p-5)-p}{2n}},\\
& y_2=\left( \frac{n}{q+n} \right) r_2^{\frac{q}{n}+1}. \\
\end{split}
\end{equation}
The solution corresponds to
\beqa
&&ds_{n+2}^2=e^{-2\sigma_1-2\sigma_2}ds^2_n+e^{-2 \sigma_2} dy_1^2 +e^{-2 \sigma_1} dy_2^2,\\
&& \phi_1=- \sqrt{\frac{n(9-p)}{9+n(p-5)-p}} \log y_1,\\
&&\phi_2= - \sqrt{\frac{qn}{q+n}}\log y_2,\\
&& \sigma_1=- \frac{(9-p)}{9+n(p-5)-p}\log y_1, \\
&& \sigma_2=- \frac{q}{q+n}\log y_2.
\eeqa
Hence the resulting critical exponents are
\beqa
\delta_1= 2  \sqrt{\frac{9+n(p-5)-p}{n(9-p)}}\quad ,\quad \delta_2=2 \sqrt{\frac{q+n}{qn}}.
\eeqa

\subsection{One general ETW brane}
\label{sec:one-general-ETW}

In this section we present  a generalization of the previous sections, and consider the intersection of a bubble of nothing ETW brane, corresponding to a shrinking $\IS^p$, with a completely general ETW brane characterized by a critical exponent $\delta$. The solution is actually very simple. We start with a theory in $d=n+p+2$ dimensions and action
\begin{equation}
    S_d = \int d^d x \sqrt{-g} \left[ \frac{1}{2} R - \frac{1}{2} \left( \partial \varphi \right)^2 - V(\varphi) \right]
    \label{daction_1scalar_potential}
\end{equation}
We consider a general codimension-1 local ETW brane solution as in (\ref{dw-ansatz}), (\ref{local-codim1-etw}), sliced along an angular $\IS^p$ on its worldvolume dimensions, namely
\begin{equation}
\begin{split}
    & ds^2_d = e^{-2 \sigma (y)} \,[\,
    ds_{n}^2+dr^2+r^2d\Omega_p^2 
    \,]\,+dy^2 \\
    & \varphi (y) \simeq - \frac{2}{\delta} \log y \quad ,\quad \sigma (y) \simeq - \frac{4}{(d-2) \delta^2} \log y, \\
\end{split}
\label{ETW-meets-BoN}
\end{equation}
The potential in the regime near the ETW brane is of the exponential form (\ref{codim1-potential}):
\begin{equation}
    V (\varphi) = -ac e^{\delta \varphi}.
\end{equation}
Taking the coordinate $r$ to parametrize a coordinate along with the $\IS^p$ varies, this realizes the intersection of ETW branes of interest.

In order to describe it from the perspective of the $(n+2)$-dimensional action after reduction along the $\IS^p$, we take the compactification ansatz
\begin{equation}
    ds^2_d = e^{\alpha \rho} ds^2_{n+2} + e^{-\beta \rho} d \Omega^2_p,
    \label{ansatz_second_compactification}
\end{equation}
The coefficients $\alpha$, $\beta$ are fixed by the Einstein frame condition in $(n+2)$-dimensional action and the normalization of the scalar $\rho$. We require
\beqa
n\alpha = p\beta \quad ,\quad \beta^2=\frac{4n}{p(p+n)}.
\eeqa
The resulting $(n+2)$-dimensional action is
\begin{equation}
\begin{split}
     S_{n+2} &=  \frac{1}{2} \int  d^{n+2}x \sqrt{-g_{n+2}} \Big\lbrace  R_{n+2} - \left( \partial \varphi \right)^2- \left( \partial \rho \right)^2+ \frac{p(p-1)}{R^2_p} e^{2 \sqrt{\frac{n+p}{np}}  \rho } + \\ 
    &\left. -2 V(\varphi) e^{2 \sqrt{\frac{p}{n(n+p)}}  \rho  } \right\rbrace, \\
\end{split}
\label{reduced_n+2_action}
\end{equation}
where $\varphi$ is the scalar associated to the generic ETW brane in $d-$dimensions and $\rho$ is the breathing mode of the $\IS^p$ compactification. In order to get an action in the form (\ref{action_2scalars}), we redefine the fields via:
\beqa
\varphi =  \delta \sqrt{\frac{n(n+p) } {4p+n(n+p) \delta^2} } \,\phi_1 \quad ,\quad 
         \rho = \phi_2 +2 \sqrt{\frac{p}{4p+n(n+p) \delta^2}}\, \phi_1 \, ,
\eeqa
and the (\ref{reduced_n+2_action}) becomes:
\begin{equation}
\begin{split}
    & S_{n+2} =  \frac{1}{2} \int  d^{n+2}x \sqrt{-g_{n+2}} \Big\lbrace  R_{n+2} - \left( \partial \phi_1 \right)^2- \left( \partial \phi_2 \right)^2 - \alpha \partial_{\rho} \phi_1 \partial^{\rho} \phi_2 + \\
    & \left. + \frac{p(p-1)}{R^2_p} e^{2 \sqrt{\frac{n+p}{np}}  \left( \phi_2 -2 \sqrt{\frac{p}{4p+n(n+p) \delta^2}} \phi_1 \right) } +2 a c e^{2 \sqrt{\frac{p}{n(n+p)}} \phi_2 + \sqrt{\frac{4p+n(n+p) \delta^2}{n(n+p)}} \phi_1 }   \right\rbrace,  \\
\end{split}
\end{equation}
with
\begin{equation}
    \alpha = 2 \sqrt{ \frac{p}{4p +n(n+p) \delta^2}}.
\end{equation}
It is easy to check that the configuration (\ref{ETW-meets-BoN}) is a local intersecting ETW brane solution (\ref{local-intersecting-etws}). Performing the change of variables
\beqa 
y_1=\left( \frac{n(n+p) \delta^2}{4p+n(n+p)\delta^2} \right) y^{\frac{4p}{n(n+p)\delta^2}+1}\quad ,\quad y_2= \left( \frac{n}{p+n}\right) r^{\frac{p}{n}+1}
\eeqa
we obtain
\beqa
ds_{n+2}^2&=& e^{-2\sigma_1-2\sigma_2}ds^2_n+e^{-2 \sigma_2} dy_1^2 +e^{-2 \sigma_1} dy_2^2 \nonumber\\
\sigma_1 &=& -\frac{4 (n+p)}{\delta ^2 n (n+p)+4 p} \log y_1 \quad , \quad \sigma_2= -\frac{p}{n+p} \log y_2 \nonumber \\
\phi_1&=&- 2 \sqrt{\frac{n(n+p)}{4p+n(n+p) \delta^2}} \log y_1 \quad ,\quad \phi_2=- \sqrt{\frac{np}{n+p}} \log y_2
\eeqa
This corresponds to the critical exponents
\beqa
    \delta_1 = \sqrt{ \delta^2+\frac{4p}{n(n+p)}}
\quad ,\quad
    \delta_2 = 2 \sqrt{\frac{n+p}{np}}.
\eeqa
Note that $\delta_2$ nicely agrees with the value obtained in section \ref{sec:sp1sp2} c.f. (\ref{delta-sp1sp2}). We can also recover examples of previous sections for different choices of $\delta$. For instance, for $\delta^2 = 4 \frac{n+p+q}{q(n+p)}$ we recover the case of $\IS^p \times \IS^q$ compactification studied in section \ref{sec:sp1sp2}. Also, for $\delta^2= \frac{4 (q-3)^2}{q(9-q)}$, with $q<7$, we recover the case of a D$q$-brane solution reduced along the transverse $\IS^{8-q}$ times an $\IS^p$ along the brane worldvolume, as studied in section \ref{sec:dbrane} (with reversed labels $p$, $q$). 

We hope that these examples suffice to illustrate that our class of solutions includes many physically relevant cases of ETW branes.

\section{Swampland applications}
\label{sec:swampland}

In this section we discuss the interplay of our solutions with various swampland constraints. We mainly focus on the cobordism conjecture and the distance conjecture, but mention others along the way.

\subsection{Cobordism conjecture}
\label{sec:swamp-cobordism}

Since ETW branes are motivated by the Cobordism Conjecture, there are several interesting interpretations of our intersecting ETW brane solutions from this perspective.

\medskip

\subsubsection{The end of the world for end of the world branes}

Dynamical cobordisms arise in the exploration of the Cobordism Conjecture \cite{McNamara:2019rup} beyond its merely topological avatar. They provide explicit effective field theory descriptions of cobordisms to nothing, including the possibly necessary defects to remove any non-trivial cobordism classes or other charges in the configuration. 

From the perspective of the effective theory, the microscopic structure of the cobordism defect remains mostly shrouded in the mist of the UV completion, but there may be features of the ETW brane worldvolume dynamics amenable to the effective theory approach. One way to probe them is to let the ETW brane interact with other defects of the theory. In the familiar context of string theory branes, the worldvolume gauge field content in a brane can be read from the objects able to end on it \cite{Strominger:1995ac}, and some such BIon configurations are accessible in supergravity \cite{Gomberoff:1999ps}. In the context of cobordism defects, for instance \cite{Dierigl:2023jdp} obtained the type IIB R7-brane worldvolume theory by the characterization of branes allowed to end on its worldvolume. Namely, cobordism defects must be able to explain not just the end of bulk spacetime, but also the disappearance of the possible defects present in the bulk theory.

Intersecting ETW brane configurations can be regarded as a radical case of this last idea: Cobordism defects must be able to explain not just the end of bulk spacetime, but also of bulk configurations bounded by other pre-existing cobordism defects. From the effective theory perspective, when the bulk theory is bounded by a first ETW brane (ETW$_1$), the resulting configuration must admit being bounded by a second ETW brane (ETW$_2$), with the bulk ending on the ETW$_2$ brane and its boundary ETW$_1$ brane ending on its intersection with the ETW$_2$ brane, see Figure \ref{fig:ETW-of-ETW}a. Obviously there is the converse picture, in which the ETW$_1$ brane provides a boundary of the configuration given by the bulk theory ending on the ETW$_2$ image. Hence our intersecting ETW brane solutions explain the end of the world for end of the world branes.

\medskip

\subsubsection{Interpolating Domain Walls between ETW branes}

There is another related but complementary interpretation of intersecting brane configurations from the Cobordism Conjecture perspective, focusing on its implication that in theories of quantum gravity any two configurations can be connected by some domain wall. Hence we may consider two configurations given by the bulk theory bounded by the ETW$_1$ and the {\em same} bulk theory bounded by a different ETW$_2$ brane. We can now consider the interpolating configuration across which the bulk theory is unchanged but the ETW$_1$-brane turns into the ETW$_2$ brane. This can be regarded as a configuration of two intersecting ETW branes in the limit where their angle is close to $\pi$, with the intersection playing the role of interpolating domain wall between the ETW brane boundaries, see Figure \ref{fig:ETW-of-ETW}b. 

ETW branes at general angle $\theta$ are explicitly constructed in appendix \ref{sec:angles}. The limit $\theta\to \pi$  is singular in that description, but only because the original coordinate system becomes degenerate. 
It would be interesting to extract the resulting limit configuration, or build its solution from scratch, but we refrain from doing so in the hope that the conceptual picture is clear enough.

The above links with the cobordism conjecture relate to the structure of spacetime and its boundaries (and boundaries of their boundaries). The realization of these configurations via dynamical cobordisms with scalars blowing up at the ETW branes  furthermore leads to an in interesting map between spacetime physics and field space, to which we turn next.

\subsection{Distance Conjecture}
\label{sec:swamp-distance}

Our solutions are spacetime-dependent configurations probing infinite distance limits in field space at finite spacetime distance, and therefore have a direct interplay with the Distance Conjecture \cite{Ooguri:2006in}. We explore various such connections in this section.

\subsubsection{General Distance Conjecture}

The Distance Conjecture \cite{Ooguri:2006in} states that effective field theories break down along geodesic paths extending to infinite distance limits in moduli space due to the appearance of an infinite tower of states at a scale $m$ falling exponentially with the field theory distance
\beqa
m\sim e^{-\lambda {\cal D}}
\label{distance-conjecture}
\eeqa
with some ${\cal O}(1)$ coefficient $\lambda$ (with $\lambda\geq \frac{1}{\sqrt{d-2}}$ according to the sharpened distance conjecture \cite{Etheredge:2022opl}).

In spacetime-dependent configurations probing infinite field space distances at finite spacetime distance, such as ETW branes \cite{Angius:2022aeq,Blumenhagen:2022mqw,Blumenhagen:2023abk}, there are interesting relations between the field space distance and the spacetime distance, c.f. (\ref{universal-scalings-codim1}). This allows to provide a spacetime version of the Distance Conjecture which expresses the falloff of the cutoff scale along some path in spacetime. 

In our intersecting ETW brane setup, we have found similar scaling relations between the field space distance and the spacetime distance (\ref{scaling-codim2}), controlled by the path-dependent coefficient $\delta_{\rm int}$. Combining this expression with (\ref{distance-conjecture}), we obtain
\beqa
m\sim \Delta^{\frac{2\lambda}{\delta_{\rm int}}}
\eeqa

In spacetime-dependent solutions, i.e. beyond the adiabatic approximation, the interpretation of $m$ is not necessarily the appearance of an infinite tower \cite{Buratti:2018xjt}; it instead indicates the scale at which new UV physics kicks in. This played an important role in the context of small black holes, where it fixes the size of the smallest possible black hole in the effective theory \cite{Hamada:2021bbz,Calderon-Infante:2023uhz}. It would be interesting to explore similar mechanisms for ETW branes.

\subsubsection{The Convex Hull Distance Conjecture}

In theories with several scalar fields, there are different infinite distance limits, which probe the existence of different UV towers. Hence, the cutoff along general infinite distance path in field space can be sensitive to multiple individual towers. 
A general recipe to obtain the exponential falloff along such a general path is provided by the Convex Hull Distance Conjecture \cite{Calderon-Infante:2020dhm} (see also \cite{Etheredge:2023odp}), as follows.

Consider all the towers of the theory corresponding to all possible infinite distance limits, and denote by $m(\phi^i)$ the moduli-dependent mass scale of each of these towers. Let us introduce the scalar charge to mass ratios
\beqa
\zeta_i \equiv \partial_i \log m .
\eeqa
Consider a trajectory $\phi^i(t)$ going to some infinite distance limit, and define the (normalized) tangent vector
\beqa
\tau^i \equiv \frac{\phi'{}^i}{||\phi'||}\quad ,\quad {\rm with}\; \phi'{}^i=\frac{d\phi^i}{dt}
\eeqa
Along this trajectory the tower scale falls off as in (\ref{distance-conjecture}) with 
\beqa
\lambda = -\zeta_i \,\tau^i =- \,\frac{\zeta_i\phi'{}^i}{||\phi'||}.
\label{rate}
\eeqa
Notice that there is a dependence on the moduli space metric in the normalization of the tangent vector. 

Introducing the tangent frame $e_i^a$ in field space $G_{ij}=\delta_{ab}e_i^a e_j^b$, and its inverse $e_a^i$, this can be recast in terms of vector scalar products 
\beqa
\lambda=-\zeta^a\tau_a\quad ,\quad {\rm with}\; \zeta^a=e_a^i \zeta_i\; ,\; \tau_a = e_a^i \tau_i.
\eeqa
Combining all towers, the falloff rate is controlled by the convex hull defined by the scalar charge to mass ratios for all towers in the theory \cite{Calderon-Infante:2020dhm}. 

We have shown that theories with several scalars admit intersecting ETW brane solutions, and that different spacetime paths approaching the intersection define different field space paths traversing infinite distance. Formally, one can regard the scalar profiles $\phi^i(y^\mu)$ in our solution\footnote{We momentarily change to upper indices for fields and spacetime coordinates in order to match usual mathematical conventions in the following argument.} as defining an embedding of two spacetime dimensions into the scalar field space. This allows to define pullbacks of moduli space quantities onto the spacetime dimensions, and formulate a spacetime avatar of the Convex Hull Distance Conjecture.

Indeed, the pullback onto spacetime of the scalar charge to mass ratio is
\beqa
\zeta_\mu=\partial_\mu \log m (\phi^i(x^\mu))=\partial_\mu \phi^i\, \zeta_i
\eeqa
Also, for a path in spacetime $x^\mu(\lambda)$, with (unnormalized) tangent vector
\beqa
v^\mu=x'{}^\mu(\lambda)\quad ,\quad {\rm with}\; x'{}^\mu=\frac{dx^\mu}{d\lambda}\, ,
\eeqa 
we get a path $\phi^i(x^\mu(\lambda))$ in field space, with (unnormalized) tangent vector 
\beqa
\frac d{d\lambda} \phi^i(x^\mu(\lambda))=\partial_\mu \phi^i\, x'{}^\mu=\partial_\mu \phi^i\,v^\mu
\eeqa
We then have the spacetime version of  the numerator of (\ref{rate}) 
\beqa
\zeta_i\, \phi'{}^i\,=\,\zeta_i\partial_\mu \phi^i\,  v^\mu=\zeta_\mu v^\mu
\eeqa
The Convex Hull criterion requires using normalized tangent vectors in field space. Using the scalar field metric $G_{ij}$, we have
\beqa
||\phi' ||=(G_{ij} \phi'{}^i\phi'{}^j)^{\frac 12}=(G_{ij}\partial_\mu \phi^i\partial_\nu\phi^j x'{}^\mu x'{}^\nu)^{\frac 12}=(h_{\mu\nu} v^\mu v^\nu)^{\frac 12}\, ,
\eeqa
namely, the norm of $v^\mu$ but computed with the induced metric 
\beqa
h_{\mu\nu}=G_{ij}\partial_\mu \phi^i\partial_\nu\phi^j
\eeqa
Hence one can formulate the Distance Conjecture as a statement in spacetime in terms of the metric $h_{\mu\nu}$. Note that this is actually different from the spacetime metric $g_{\mu\nu}$. In particular, the field space metric contains mixed terms, whereas the actual spacetime metric is diagonal. 
It would be interesting to discuss dynamical properties in spacetime of this induced metric from the field space.

\subsubsection{The infinite distance pattern}

The above ideas can be easily extended to other swampland criteria. For instance, in \cite{Castellano:2023stg,Castellano:2023jjt}, an interesting pattern was proposed to hold at infinite distance limits
(see also \cite{Rudelius:2023spc} for proposals in the interior) between the tower scale $m(\phi^i)$ and the species scale (the effective cutoff scale of quantum gravity \cite{Dvali:2007hz,Dvali:2008ec,Dvali:2009ks,Dvali:2010vm,Dvali:2012uq}) $\Lambda_s(\phi^i)$, which in general is moduli-dependent. Specifically, they are claimed to satisfy
\beqa
G^{ij} \,\partial_i \log m\, \partial_j \log \Lambda_s=\frac{1}{d-2}
\eeqa
In the context of our solutions, there is a spacetime version of this condition using the (inverse) induced metric
\beqa
h^{\mu\nu} \,\partial_\mu \log m (\phi^i(y^\mu))\, \partial_\nu \log \Lambda_s(\phi^i(y^\mu))=\frac{1}{d-2}
\eeqa

\medskip

Note that the species scale and its relation to the distance conjecture has already been studied from the perspective of the link between spacetime and field space structures for codimension-1 ETW brane solutions in  \cite{Calderon-Infante:2023ler}. We expect that similarly exciting ideas may arise in the codimension-2 case of intersecting ETW branes. We leave these explorations as well as links to other swampland conjectures for future work. 

\section{Conclusions}
\label{sec:conclu}

The exploration of infinite distance limits has led to the construction of diverse defects, such as ETW branes, small black holes or 4d EFT strings, defined by the fact that scalars reach infinite field theory distance at their cores. It is natural to ask about the interplay of such objects, and their use to explore the network of infinite field theory distance limits, i.e. their different components and their intersections.

In this paper we have initiated this exploration by constructing explicit solutions describing intersecting ETW branes in theories with multiple scalars. The configurations behave as the superposition of two codimension-1 ETW branes, and display interesting path-dependent scaling properties along trajectories approaching the codimension-2 intersection.  We have explored the interplay of these solutions with swampland conjectures, and in particular with the convex hull description of the Distance Conjecture in theories with several scalars. Finally, we have explicitly shown that many interesting systems correspond to solutions within our class, including intersections of bubbles of nothing and several generalizations thereof.

Some of the interesting questions opened up by our work are:

$\bullet$ Our solutions can be regarded as a mere superposition of ETW branes, in the sense that their source terms are localized on the individual codimension-1 ETW branes. It would be interesting to describe more general intersections supported also by codimension-2 source terms. In this respect, it would be interesting to connect with the codimension-2 objects in \cite{Blumenhagen:2022mqw,Blumenhagen:2023abk}.

$\bullet$ There are localized D-brane solutions in the literature (see \cite{Ortin:2015hya} for review and references), which upon suitable reductions along transverse space may be rephrased as intersections of codimension-1 ETW branes. However, the reduction involves directions along which there are no isometries and it is likely it leads to solutions not describable within our ansatz. It would be interesting to study these examples as a tool to generate more general solutions, in particular including examples with non-trivial degrees of freedom at the intersection of ETW branes.

$\bullet$  In our solutions the scalars diverging at the ETW branes had a non-trivial mixed kinetic term. It would be interesting to describe intersecting configurations for decoupled scalar fields, in particular to better connect with CY moduli spaces near infinite distance limits \cite{Grimm:2018ohb,Grimm:2018cpv,Corvilain:2018lgw}. This will be addressed in \cite{Angius:2024}.

$\bullet$ We have focused on ETW branes and their interplay via intersections. More generally, it would be interesting to understand the interplay of other defects defined by scalars running off to infinity in field space. For instance, the crossing of two EFT strings may lead to the creation of new strings, unveiling non-abelian structures at infinity in moduli space, in the spirit of \cite{Berasaluce-Gonzalez:2012abm}. 

We hope to come back to these and other interesting questions in future work.


\section*{Acknowledgments}

We are pleased to thank Bruno Bento, José Calderón-Infante, Matilda Delgado, Jesús Huertas, Luis Ibáñez, Christian Kneissl, Fernando Marchesano, Miguel Montero,  Irene Valenzuela for useful discussions. This work is supported through the grants CEX2020-001007-S and PID2021-123017NB-I00, funded by MCIN/AEI/10.13039/501100011033 and by ERDF A way of making Europe. The work by R.A. is supported by the grant BESST-VACUA of CSIC. 

\newpage

\appendix 

\section{Some generalizations}
\label{sec:generalizations}

In this appendix we consider several generalizations and variants of the solutions discussed in the main text.

\subsection{Intersection at angles}
\label{sec:angles}

In the main text we have restricted the intersection of ETW branes to be orthogonal in the conformally flat coordinates (\ref{conf-flat}). It is a natural generalization to consider a general off-diagonal term
\beqa
ds_{n+2}^2 &=& e^{-2\sigma_1-2\sigma_2}\,[\, ds_n^2 \,+ \,   dx_1^2\,+\,dx_2^2 \,+\, f\, dx_1dx_2 \,]\nonumber\\
&=& e^{-2\sigma_1-2\sigma_2}\, ds_n^2\,+\, e^{-2\sigma_2}\, dy_1^2\,+\,e^{-2\sigma_1}\, dy_2^2\,+\,f\,e^{-\sigma_1-\sigma_2}\, dy_1dy_2\, ,
\label{conf-flat2}
\eeqa
where $\sigma_i$ are regarded as functions of $x_i$ in the top line and of $y_i$ in the bottom one. We will look for solutions in which each scalar still runs along one coordinate $\phi_1(y_1)$, $\phi_2(y_2)$. In fact, we try and solve the equations of motion by using logarithmic profiles for $\sigma_i$ and $\phi_i$ as in (\ref{log-ansatz}), which we repeat for convenience
\beqa
&&\sigma_1=-a_1 \log y_1 + \frac{1}{2} \log c_1, \quad \quad \sigma_2=-a_2\log y_2 + \frac{1}{2} \log c_2 ,\nonumber \\
&& \phi_1=b_1\log y_1, \quad \quad \quad \quad  \ \quad \quad \phi_2=b_2\log y_2,
\label{log-ansatz2}
\eeqa
The equations of motion are satisfied if the coefficients satisfy:
\beqa
b_i^2=na_i^2\quad , \quad \alpha = 2\sqrt{a_1a_2}
\label{thesame}
\eeqa
and the potential behaves as
\beqa
&& V = V_1+ V_2 + \mathcal{O} \left( y_1^{-1-a_1}y_2^{-1-a_2} \right)\nonumber \\
 &&V_1= c_1 v_1 y_1^{-2}y_2^{-2a_2} = \frac{2 c_1 n a_1 }{ \left( f^2 c_1 c_2 -4 \right)}   \left[ (na_1+a_1-1)  \right] y_1^{-2} y_2^{-2a_2} \\
       & &V_2= c_2 v_2 y_1^{-2a_1}y_2^{-2a} = \frac{2 c_2 n a_2 }{ \left( f^2 c_1 c_2 -4 \right)}   \left[ (na_2+a_2-1)  \right] y_1^{-2a_1} y_2^{-2} \\
       & & \mathcal{O}\left( y_1^{-1-a_1}y_2^{-1-a_2} \right) = - c_1 c_2 \frac{2n^2f a_1 a_2}{\left( f^2 c_1 c_2 -4 \right)} y_1^{-1-a_1}y_2^{-1-a_2}
\eeqa
Note that the equations of motions provide us more than two terms for the potential. Assuming that $a_1<1$ and $a_2<2$, the last term is subleading respect to the previous ones which contain the information to specify the kind of intersecting ETW branes.

Moreover the coefficients are given by a generalization of (\ref{the-as}), namely
\begin{equation}
    a_i = \frac{n \pm  \sqrt{n+ 2 (n+1) v_i \left(f^2 c_1 c_2-4\right)}}{2 n (n+1)}
\end{equation}
Finally, the critical exponents are given by
\begin{equation}
    \delta_i^2 = \frac{4}{a_1 n} = \frac{8(n+1)}{n \pm \sqrt{n +2(n+1)v_i (f^2c_1 c_2 -4)}}
\end{equation}

Hence, even though (\ref{thesame}) has the same structure as (\ref{constraints-eoms}), there is a non-trivial dependence on $f$ in the potential and the coefficients of the logarithms. This implies that, given the potential of the theory, one can read off the values of $a_i$ and $v_i$ from its leading exponential behaviour, and determine the value (or values) of $f$ that provides a solution, which in general corresponds to non-orthogonal intersections.

Note that when $f^2 c_1 c_2-4=0$ the values above become singular. This corresponds to the limit where the angle between the two ETW branes is $0$ or $\pi$ and the two ETW branes overlap. It would be interesting to explore the limiting behaviour near this regime.

\subsection{Triple intersections and beyond}
\label{sec:triple}

In this section we discuss a natural generalization of the intersecting ETW branes of section \ref{sec:intersecting}, by considering triple intersections of three independent ETW branes.

We consider the following action for $(n+3)-$dimensional gravity coupled to three real scalar fields with general potential $V(\phi_1, \phi_2, \phi_3)$:
 \begin{equation}
     S_{n+3} = \int d^{n+3} \sqrt{-g} \left[ \frac{1}{2} R - \frac{1}{2} \sum_{i=1}^3 \left( \partial \phi_i \right)^2 - \frac{1}{2} \sum_{i \neq j} \alpha_{ij} \partial_{\rho} \phi_i \partial^{\rho} \phi_j - V(\phi_1, \phi_2, \phi_3) \right].
 \end{equation}
We consider the following ansatz for the metric:
\begin{equation}
    ds^2_{n+3} = e^{2A(y_1,y_2,y_3)} ds^2_n + e^{2B(y_1,y_2,y_3)}dy_1^2 +e^{2C(y_1,y_2,y_3)}dy_2^2+ e^{2D(y_1,y_2,y_3)}dy_3^2 .
\end{equation}
The equations of motion admit solutions with each scalar $\phi_i$ depends only on the coordinate $y_i$, given by a simple generalization of (\ref{codim2-ansatz}), namely
\beqa
ds_{n+3}^2=e^{-2\sigma_1-2\sigma_2-2\sigma_3}ds^2_n+e^{-2\sigma_2-2\sigma_3}dy_1^2+e^{-2\sigma_1-2\sigma_3}dy_2^2+e^{-2\sigma_1-2\sigma_2} dy_3^2 ,
\eeqa
while the different functions are locally of the form
\beqa
\phi_i=b_i \log y_i\quad , \quad \sigma_i=-a_i\log y_i + \frac{1}{2} \log c_i
\eeqa
and the scalar potential splits into three terms $V=V_1+V_2+V_3$ encoding the dominant terms as the different scalars go off to infinity, with
\beqa
V_i=- c_i v_i y_i^{-2} y_j^{-2 a_j} y_k^{-2a_k}\quad , \; i\neq j\neq k
\eeqa
The parameters of the solution are related by
\beqa
b_i^2=(n+1)a_i\quad , \quad a_i=\frac{1\pm\sqrt{1+8v_i\frac{n+2}{n+1}}}{2(n+2)}\quad ,\quad \alpha_{ij}=(a_ia_j)^{-\frac 12} .
\eeqa
Hence the critical exponents are
\beqa
\delta_i^2= \frac{8 (n+2)}{(n+1) \pm (n+1) \sqrt{1+8v_i \frac{n+2}{n+1}}} .
\eeqa
in terms of which the analogue of (\ref{local-intersecting-etws}) is
\beqa
V_i&=&- c_i v_i e^{\delta_i\phi_i} e^{a_j\delta_j\phi_j} e^{a_k\delta_k\phi_k} \quad , \; i\neq j\neq k \nonumber\\
\phi_i&=&-\frac{2}{\delta_i}\log y_i\quad ,\quad 
\sigma_i= -\frac{4}{(n+1) \delta_i^2} \log y_i\, .
\label{local-intersecting-etws-triple}
\eeqa
It is clear that one can generalize to even higher-codimensional intersections. Triple or higher intersections can be useful to further understand intersection of loci corresponding to multiple infinite distance limits.

\subsection{ETW brane configurations with a single scalar field}
\label{sec:single}

One may wonder to what extent we need two scalars to achieve intersecting ETW brane configurations. In this appendix we consider candidates for codimension-2 intersections of ETW branes in a theory with a single scalar. We will show that the configuration is actually better described as a single recombined codimension-1 ETW brane.

We start with $(n+2)$-dimensional gravity coupled to one real scalar with general potential 
\begin{equation}
    S= \int d^{n+2}x \sqrt{-g} \left[ \frac{1}{2 }R - \frac{1}{2} \left( \partial \phi \right)^2 - V(\phi) \right]\, .
\end{equation}
We consider a codimension-2 ansatz
Considering the following ansatz for the solution:
\beqa
        && ds^2_{n+2} = e^{2 A (y_1 , y_2)}ds^2_n + e^{2B(y_1,y_2)} dy^2_1 + e^{2C(y_1,y_2)}dy^2_2 \nonumber\\
        && \phi = \phi(y_1,y_2), 
        \label{ansatz-single}
    \eeqa
and we solve the equations of motion with a set of logarithmic profiles for the fields
\beqa
        &&A = a_1 \log y_1 + a_2 \log y_2  \quad ,\quad B= b_2 \log y_2  \quad ,\quad
         C = c_1 \log y_1 \nonumber\\
        && \phi = d_1 \log y_1+ d_2 \log y_2
        \label{logs-single}
\eeqa
We get the conditions
\beqa
& (a_2-b_2)(na_2+b_2-c_2)-a_2+b_2=0 \quad & ,\quad
 (a_1-c_1)(na_1+c_1-b_2)-a_1+c_1=0 \nonumber \\
& d_1^2 = c_1(a_1-c_1) +(n-1)a_1+c_1 \quad\quad &,\quad  d_2^2 = b_2(a_2-b_2) +(n-1)a_2+b_2\nonumber\\
& d_1^2 = \frac{1}{2} na_1 \left(1 + \frac{d_1}{d_2} b_2 \right) \quad &,\quad
 \d_2^2 = \frac{1}{2} na_2 \left(1 + \frac{d_2}{d_1} c_1 \right) \nonumber\\
& V_{1} = - \frac{n}{2} a_1 \left( na_1 +c_1 -1\right)y_1^{-2} y_2^{-2b_2} \quad &,\quad
 V_{2} = - \frac{n}{2} a_2 \left( na_2 +b_2 -1\right)y_1^{-2 c_1} y_2^{-2} \nonumber 
\eeqa
This system is solved by the following choice of the parameters:
\beqa
         d_1 = d_2 = - \sqrt{n} \quad , \quad a_1=c_1=a_2=b_2=1
         \label{params-single}
    \eeqa
The scaling relations are now satisfied with the critical exponent:
\begin{equation}
    \delta= \frac{2}{\sqrt{n}}\, ,
    \label{delta-single}
\end{equation}
independently of the path. Indeed, the computation is basically identical to that of section \ref{sec:scalings} with a single field, so that $\delta_1=\delta_2\equiv \delta$. The analogue of (\ref{delta-int}) is 
\beqa
\delta_{\rm int}=  (r_i+1)\, \delta
\label{delta-int}
\eeqa
with $r_i$ as in (\ref{the-rs}). Actually, using a parametrization satisfyin $\gamma_1+\gamma_2=1$, the latter are $r_i=0$, and hence $\delta_{\rm int}=\delta$ independently of the path.

The interpretation is that the configuration, rather than the intersection of two individual ETW branes, describes a singular limit of a single recombined ETW brane. This is also motivated by the fact that the solution (\ref{ansatz-single}), (\ref{logs-single}), with parameters (\ref{params-single}), is
\beqa
ds_{n+1}^2= y_1^2y_2^2 ds_n^2 + y_2^2 dy_1^2 + y_1^2 dy_2^2\quad ,\quad \phi=-\sqrt{n} \log (y_1 y_2)
\eeqa
Performing a change of coordinates
\beqa
y=y_1y_2\quad ,\quad x=\log(y_1/y_2)
\eeqa
we see that the scalar profile varies non-trivially only along $y$. In fact, the full solution reads
\beqa
ds_{n+1}^2= y^2 (\, ds_n^2 + 2dx^2\,) +  2dy^2\quad ,\quad \phi=-\sqrt{n} \log (y)
\eeqa
Reabsorbing the factors of 2 via simple redefinitions, this corresponds to a codimension-1 ETW brane solution of type (\ref{dw-ansatz}), (\ref{local-codim1-etw})), for $\delta$ given in (\ref{delta-single}). Note the interesting way in which the coordinate $x$, along which the scalar is constant, becomes a coordinate along the codimension-1 ETW brane, by combining with the $n$ original ones.

Actually, this phenomenon of recombination of ETW branes was already proposed in \cite{Angius:2022mgh} in the particular context of ETW branes in supercritical bosonic strings with light-like tachyon condensation. The system contains two ETW branes with precisely the critical exponent (\ref{delta-single}). One is spacelike and corresponds to the dilaton growing towards infinitely strong coupling at a point in a finite past time in the Einstein frame; the second is lightlike and corresponds to closed string tachyon condensation. Both ETW branes meet in a codimension-2 $(D-2)$-dimensional locus. Although the ETW branes seem to involve two different scalars, the dilaton and the tachyon, it was shown that they mix together into a single combination. This motivated the proposal that the two ETW branes should be regarded as a single recombined one, so that the beginning of time can be thought of as a strong coupling avatar of closed string tachyon condensation.
It is very satisfactory that our general analysis here provides extra support for this picture.

\bibliographystyle{JHEP}
\bibliography{mybib}

\end{document}